\documentclass{nature}

\usepackage{upgreek}
\usepackage{amssymb}
\usepackage{times}
\usepackage{graphicx}
\usepackage{bm}
\usepackage{color}
\usepackage{soul}
\newcommand{\CII}{[C\,\textsc{ii}]}
\newcommand{\CIV}{C\,\textsc{iv}}
\newcommand{\OI}{O\,\textsc{i}}

\newcommand{\msolyr}{M_\odot \, \rm yr^{-1}}
\newcommand{\kms}{km~s$^{-1}$}

\def\ltsima{$\; \buildrel < \over \sim \;$}
\def\simlt{$\lower.5ex\hbox{\ltsima}$}
\def\gtsima{$\; \buildrel > \over \sim \;$}
\def\simgt{$\lower.5ex\hbox{\gtsima}$}
\def\arcsec{$''~$}

\newcommand{\msun}{M_{\odot}}
\newcommand{\smyr} {\msun \mbox{ yr}^{-1}}

\makeatletter
\let\saved@includegraphics\includegraphics
\AtBeginDocument{\let\includegraphics\saved@includegraphics}
\renewenvironment*{figure}{\@float{figure}}{\end@float}
\makeatother

\title{A \CII\ 158$\mu$m Emitter Associated with an \OI\ Absorber at the End of the Reionization Epoch}

\author{Yunjing Wu$^{1}$, Zheng Cai$^{1,2,3*}$, Marcel Neeleman$^{4}$, Kristian Finlator$^{5,6}$, Shiwu Zhang$^{1}$,  J. Xavier Prochaska$^{7,8}$, Ran Wang$^{9}$, Bjorn H.\,C. Emonts$^{10}$, Xiaohui Fan$^{11}$, Laura C. Keating$^{12}$, Feige Wang$^{11,16}$, Jinyi Yang$^{11,17}$, Joseph F. Hennawi$^{13}$, Junxian Wang$^{14,15}$}

\begin{document}

\maketitle

\begin{affiliations}
\item Department of Astronomy, Tsinghua University, Beijing 100084, China
\item Center for Astronomy Technology, Tsinghua University, Beijing 100084, China
\item Peng Cheng Laboratory, Shenzhen, China
\item Max-Planck-Institut f\"{u}r Astronomie, K\"{o}nigstuhl 17, D-69117, Heidelberg, Germany
\item New Mexico State University, Las Cruces, 88003 NM, USA
\item Cosmic Dawn Center (DAWN), Niels Bohr Institute, University of Copenhagen / DTU-Space, Technical University of Denmark
\item Department of Astronomy \& Astrophysics, UCO/Lick Observatory, University of California, 1156 High Street, Santa Cruz, CA 95064, USA
\item Kavli Institute for the Physics and Mathematics of the Universe (Kavli IPMU), The University of Tokyo, 5-1-5 Kashiwanoha, Kashiwa, 277-8583, Japan
\item Kavli Institute for Astronomy and Astrophysics, Peking University, Beijing 100871, China
\item National Radio Astronomy Observatory, 520 Edgemont Road, Charlottesville, VA 22903
\item Steward Observatory, University of Arizona, 933 North Cherry Avenue, Tucson, AZ 85721, USA
\item Leibniz-Institut f\"ur Astrophysik Potsdam, An der Sternwarte 16, 14482 Potsdam, Germany
\item Department of Physics, University of California, Santa Barbara, CA 93106-9530, USA
\item CAS Key Laboratory for Research in Galaxies and Cosmology, Department of Astronomy, University of Science and Technology of China, Hefei, Anhui 230026, China
\item School of Astronomy and Space Science, University of Science and Technology of China, Hefei 230026, China
\item Hubble Fellow
\item Strittmatter Fellow
\end{affiliations}

\begin{abstract}
The physical and chemical properties of the circumgalactic medium (CGM) at $z\gtrsim6$ have  been studied successfully through the absorption in the spectra of  background Quasi-Stellar Objects (QSOs) \cite{Codoreanu2017, Becker2019, Cooper2019}. One of the most crucial questions is to investigate the nature and location of the source galaxies that give rise to these early metal absorbers\cite{Cai2017, Bielby2020, Diaz2020}.  Theoretical models suggest that  momentum-driven outflows from typical star-forming galaxies can eject metals into the CGM and the intergalactic medium (IGM) at z=5--6 \cite{Oppenheimer2009, Finlator2013, Finlator2020}. Deep, dedicated surveys have searched for Ly$\alpha$ emission associated with strong CIV absorbers at $z\approx 6$, but only a few Ly$\alpha$ emitter candidates have been detected. Interpreting these detections is moreover ambiguous because Ly$\alpha$ is a resonant line\cite{Dijkstra2006, Zheng2010, Hayes2010}, raising the need for complementary techniques for detecting absorbers' host galaxies. Here, using Atacama Large Millimeter Array (ALMA), we report a \CII\ 158$\mu$m emitter associated with a strong low-ionization absorber, \OI, at $z=5.978$. The projected impact parameter between \OI\ and [CII] emitter is 20.0 kpc. The measured \CII\ luminosity is $7.0\times 10^7$ solar luminosities. Further analysis indicates that strong OI absorbers may reside in the circumgalactic medium of massive halos one to two orders of magnitude more massive than expected values \cite{Finlator2013,Keating2016}. 
\end{abstract}

Metal absorption systems (e.g., \OI, \CIV, and Mg\,\textsc{ii}) are powerful probes of the enrichment of the high-redshift intergalactic medium (IGM) \cite{Codoreanu2017,Becker2019, Cooper2019}. Because the first excitation energies of \OI\ $\lambda1302$ and neutral hydrogen are almost identical, \OI\ is considered one of the best indicators to trace the metal enrichment and the neutral IGM close to the reionization epoch \cite{Oh2002, Furlanetto2003}. Simulations suggest that  at $z=5-6$, feedback from star-forming galaxies can transport metals such as oxygen to 50 proper kpc ($\approx9''$) [ref. \cite{Oppenheimer2009, Finlator2013, Finlator2020}]. Measurements of the impact parameters and the star formation rates (SFRs) of the source galaxies can directly constrain the efficiency of galactic winds in transporting metals and then test different feedback models $^{14}$. We select one of the strongest -UV \OI$~\lambda1302$ absorber at $z=5.978$ toward QSO~J2054$-$0005 (at $z=6.04$) as our preliminary target. The absorber has a rest-frame equivalent width (REW) of 0.12\AA\ [ref.\cite{Becker2011}], corresponding to the best-fit column density of $10^{14.2}$ cm$^{-2}$ in the small optical depths regime (eq. (8) in ref.\cite{Keating2016}).

The \CII\ moment-0 map shown in Fig.~1a is constructed by collapsing the emission line channels from the \CII\ data cube. At the redshift of the \OI\ absorber, an emission line is detected at 4.3-$\sigma$ significance which we interpret as \CII\ emission from a galaxy at the \OI\ absorber's redshift. We refer this \CII\ emitter as \CII2054 in the following.
{\color{black} To check the reliability of this detection, we searched for signals in the whole datacube firstly. No sources having either a signal higher than 4-$\sigma$ level or lower than -4$-\sigma$ (Extended Data Fig. 1 and 2). }
Note that there is a low probability of $10^{-6}$ for this line to be a CO interloper (Methods). 
{\color{black} Then,} we show the spectrum of \CII2054 in Fig.~1d. The yellow shaded region represents the collapsed channels. The velocity-integrated \CII\ flux density is measured to be ${0.0758}\ \pm\ 0.0177$ Jy km s$^{-1}$, corresponding to a \CII\ luminosity of $7.0\times10^7\ L_\odot$. 
We have further discussed the reliability of this detection by checking the XX/YY polarization-correlation and individual exposure maps ({\color{black}see Extended Data Fig. 3 and 4} and details in Methods). 
{\color{black} Moreover, we further applied our target selection criteria and algorithm in the ASPECS survey\cite{Uzgil2021} over 4.2 arcmin$^2$ and found that the estimated probability of \CII2054 caused by the noise fluctuation is $\approx0.6$\% (See Extended Data Fig. 5 and details in Methods).}
Further, from our moment-0 image (Fig.~1a), the projected impact parameter between \CII2054 and the \OI\ absorber is 3.5\arcsec, corresponding to 20.0 proper kpc (pkpc) at $z=5.978$. Throughout this paper we use $\Omega_{\Lambda}=0.7$, $\Omega_{\rm M}=0.3$ and $H_0 = 70$~\kms~Mpc$^{-1}$.

Other than QSO~J2054, in the far-infrared dust-continuum image (Fig.~1b), we detect another five continuum targets with signal-to-noise ratio above four. The archival Hubble Space Telescope ({\it HST}) broad-band data have been used to estimate their photometric redshifts. All of these galaxies are securely identified as foreground sources which do not associate with the \OI\ absorber at $z=6$ ({\color{black} Extended Data Fig. 6 and 7} in Methods ). 

The \CII\ 158-$\mu$m fine structure line emission is the dominant cooling line in the interstellar medium, and it is almost unaffected by dust attenuation\cite{Carilli2013}. ALMA allows us to use \CII\ to probe the physical conditions of the gas in galaxies, and quantify the star formation rate (SFR) of the galaxy in the post-reionzation epoch\cite{Lagache2018}. The velocity-integrated \CII\ luminosity of \CII2054 is estimated to $L_{{\rm [CII]}} = (7.0\pm 1.7)\times10^{7}L_{\odot}$ [ref. \cite{Wang2013}], corresponding to a \CII-based SFR (${\rm SFR_{\rm{[CII]}}}$) of $6.8\pm1.7 M_{\odot}$ $\rm{yr^{-1}}$ [ref. \cite{Schaerer2020}]. No submillimeter continuum is detected at the position of \CII2054, yielding a 3-$\sigma$ upper limit of 37 $\mu$Jy. Converting this luminosity to a total infrared SFR ($\rm{SFR}_{\rm{TIR}}$) yields the upper limit $\rm{SFR}_{\rm{TIR}}< 11$ $\smyr$ [ref. \cite{Murphy2011}]. To estimate the halo mass directly from the \CII luminosity, we adopt a relation between $L_{\rm{[CII]}}$ and halo mass at $z\approx6$ {\color{black}[ref. \cite{Leung2020}]}, by assuming that \CII2054 resides in the galaxy main sequence. Then, the derived halo mass is $4.1\times10^{11}\ M_\odot$, with 1-$\sigma$ scattering of 0.5 dex. We also estimate the halo mass by converting the SFR to the stellar mass and then to the halo mass. The two methods of halo mass estimates yield consistent results within the 1-$\sigma$ level (see methods). 

To constrain models of IGM metal enrichment at the end of reionization, we compare our observed impact parameter against predictions from the Illustris, Sherwood, HVEL, and FAST cosmological hydrodynamic simulations\cite{Keating2016, Vogelsberger2014a, Bird2014, Bolton2017}. As mentioned above, the \OI\ absorber has a REW of 0.12\AA, and the measured projected impact parameter is 20.0 pkpc (Fig. 1c). Broadly, simulations predict that metal line absorbers with REW of $0.12 \pm 0.05$ \AA\ arise at impact parameters of 5--30 pkpc, bracketing our measurement. Specifically, in Fig.~2 (left panel), we show the comparison of our data with the Illustris simulation\cite{Keating2016}. The comparisons with other simulations are also shown in {\color{black}Extended Data Fig. 8 and}  in Methods. In this panel, the red star shows our observations, while other points are from the Illustris simulation. The simulation is able to predict our observed impact parameter, and has demonstrated that galaxies similar to \CII2054 are able to enrich the IGM at distances of $\gtrsim20$ pkpc. One should also note that $11.1\%$ of the data points in the simulation have a similar or larger impact parameter than that of \CII2054. Thus, the projected distance between \OI\ and \CII2054 is larger than median value expected from the hydrodynamical simulations.

For a direct comparison of the models, we need to derive the halo mass of \CII2054 based on SFR. As we mentioned above, \CII2054 has a halo mass of $4.1\times 10^{11}\ M_\odot$, with 1-$\sigma$ uncertainty of 0.5~dex. Different simulations, including Illustris, Sherwood, HVEL, and FAST simulations, directly investigated the halo mass distribution for \OI\ absorbers with column densities of $N_{\rm OI} > 10^{14}\ {\rm cm^{-2}}$ at $z=6$. They find that \OI\ absorbers with REW in the range $0.12 \pm 0.05$~\AA\ at $z=6$ are typically embedded in $10^{9-10}\ \rm{M_{\odot}}$ halos \cite{Finlator2013,Keating2016}. This value is one to two orders of magnitude smaller than that implied by our observations. Thus our results suggest that low-mass systems in these simulations are likely to be overly efficient at producing metal absorbers. 

A caveat to this result stems from the fact that the analysis of these simulations identified each metal absorber with its nearest halo. In reality, the nearest halo may not necessarily to be the true source of an absorber because a halo that is slightly more distant but significantly more massive may contribute more of its metals. This consideration indicates the need for an alternative framework for quantifying the galaxy-absorber relationship that can circumvent this ambiguity. 

Recently, a set of Technicolor Dawn simulations have been developed, aiming to enable a detailed study of the metal enrichment of galaxies in the reionization epoch \cite{Finlator2020}. This set of simulations incorporates realistic small-scale UV background fluctuations and reproduces a broad range of observations of reionization and early galaxy growth \cite{Finlator2020}. They simultaneously reproduce observations of the galaxy stellar mass function, the Si\,\textsc{iv} column density distribution, and the UVB amplitude at $z>5$. Hence, the galaxy populations and their metallicity of the circumgalactic medium (CGM) are realistic. 

Our discovery of \CII2054 is not expected from the Technicolor Dawn simulations. In Fig.~2 (right panel), we show how the number of galaxies that cluster about strong \OI\ absorbers at $z=6$ is predicted to vary with \CII-based SFR. The solid and dashed curves are generated by compiling catalogs of all galaxies that fall within 50 and 100 pkpc from strong \OI\ absorbers, respectively. We note that these two lines are normalized by the number of \OI\ absorbers identified within the simulation (Methods). The strong \OI\ absorbers have rest-frame equivalent widths of REW $>0.12$\AA~($N_{\rm OI}>10^{14.2}\mathrm{cm}^{-2}$). The SFR on the lower x-axis is the 100-Myr average for physically-associated galaxies, while the top x-axis converts this SFR to $L_{{\rm [CII]}}$\cite{Schaerer2020}. We plot \CII2054's SFR in this panel using red square with error bar. Consistent with previous simulations~\cite{Finlator2013, Oppenheimer2009}, Technicolor Dawn simulations predict that the characteristic luminosity of physically-associated \CII\ emitters should be roughly two orders of magnitude lower in $L_{{\rm [CII]}}$ than \CII2054; that is, far fewer than one such galaxy is predicted to lie within the area probed by our ALMA observation.

While Fig.~2 appears to suggest that our observations can be accommodated within the Illustris physical model even though it is largely unexpected in Technicolor Dawn, we note that the latter simulation's resolution is a factor of $\approx 5\times$ higher while its cosmological volume is relatively small to contain a representative population of bright galaxies $({\rm M_{\rm UV}} < -20)$, making the two simulations complementary. Given that faint galaxies with $L = 0.01 L^{*}$ are one order of magnitude more typical than massive galaxies with $L^{*}$ and that they are able to host strong OI absorbers in both models, both models predict the observational identification of a massive host to be unlikely. We further quantify the discrepency between simulations and observations by re-casting the observations as a constraint on the galaxy-absorber cross-correlation function. Based on the \CII\ luminosity function at this redshift\cite{Yan2020}, the mean space density of galaxies at $z=5.9$ with the SFR$\ge6.8\smyr$ is $1.4\times10^{-3}$ cMpc$^{-3}$. A sphere of radius 20 pkpc containing one such galaxy has a probability of $2\times10^{-5}$, corresponding to a mean overdensity of $\Delta=5\times10^4$ (Methods). Our observations therefore confirm that bright galaxies and strong absorbers correlate strongly~\cite{Diaz2020, Finlator2020}. We quantify this correlation using the galaxy-absorber cross-correlation function $\xi(r)=(r/r_0)^{-\gamma}$, which expresses the fractional excess number density of galaxies located at a distance $r$ from an absorber in terms of the correlation length $r_0$ and a power-law slope $\gamma$ (See detailed calculations of $r_0$ and $\gamma$ in Methods).

In Fig.~3, we use solid curves to show combinations of $r_0$ and $\gamma$ that predict an average of one galaxy within 20.0 pkpc of an absorber at $z=5.9$. Shaped points indicate cross-correlation functions directly predicted by Technicolor Dawn for galaxies clustered around synthetic OI absorbers with REW $\ge 0.12$\AA. As galaxies with SFR$\ge6.8\smyr$ are too rare to arise in the current simulation volume, we consider three lower SFR thresholds (see the legend). In all cases, the predicted $r_0$ is $\sim10\times$ too low to explain the observation. Moreover, while $r_0$ does increase with SFR threshold~\cite{Finlator2020}, the trend is not nearly strong enough to explain the discrepancy.

\CII2054 may be explained if its SFR were overestimated because faint galaxies are more abundant than bright ones. For example, it could be undergoing a temporary starburst. Alternatively, a strong intervening gravitational lens could boost its flux. To evaluate whether these possibilities would be sufficient to explain our observations, we use dashed and dotted curves to plot combinations of $r_0$ and $\gamma$ that would explain \CII2054 if its true SFR were ten and 100 times lower, respectively. Even in these cases, the simulation underpredicts $r_0$ by a factor of 10.

In summary, our {\color{black}} ALMA detection of a \CII\ emitter associated with a reionization-epoch \OI\ absorber is more than one to two orders of magnitude higher in \CII\ luminosity than expected from hydrodynamical simulations. Casting this observation as a constraint on the galaxy-absorber cross-correlation function suggests that the correlation length predicted by Technicolor Dawn is $\sim10\times$ too low. Our observations suggest that simulations may incorrectly identify faint galaxies as the principal repositories of observed low-ionization metal ions. This could indicate that the predicted environments of brighter galaxies are overly-ionized or that simulated faint galaxies are too efficient at creating and/or expelling metals. On the other hand, it remains possible that our ALMA observations lead us to overestimate the underlying SFR and halo mass of \CII2054. Future surveys will be needed to test conclusively whether bright galaxies are more common about strong \OI\ absorbers than expected from current theoretical models, as currently suggested by our data. Meanwhile, testing the strong prediction that many faint galaxies should exist in the vicinity of our \OI\ absorber will require a larger survey using the {\it James Webb Space Telescope (JWST) }. The three-hours {\it JWST} grism observations allow us to detect galaxies with $M_{\rm UV} \simlt -19.5$, corresponding to halomass of $10^{11}\ M_\odot$. Based on the survey depth, the mean galaxy number density\cite{Smit2012}, and the galaxy-absorber cross correlation function based on our pilot ALMA program, we expect JWST can uncover $\approx1$ host galaxies per QSO field. If we observe the field of 10 existing OI absorbers at $z\gtrsim6$ with the total integration of 30-hours, we will construct a factor of 10$\times$ larger sample, ultimately test if massive galaxies are commonly assoicated with \OI\ absorbers.

\newpage
\begin{figure}[!t]
\centering
\includegraphics[width=0.8\textwidth]{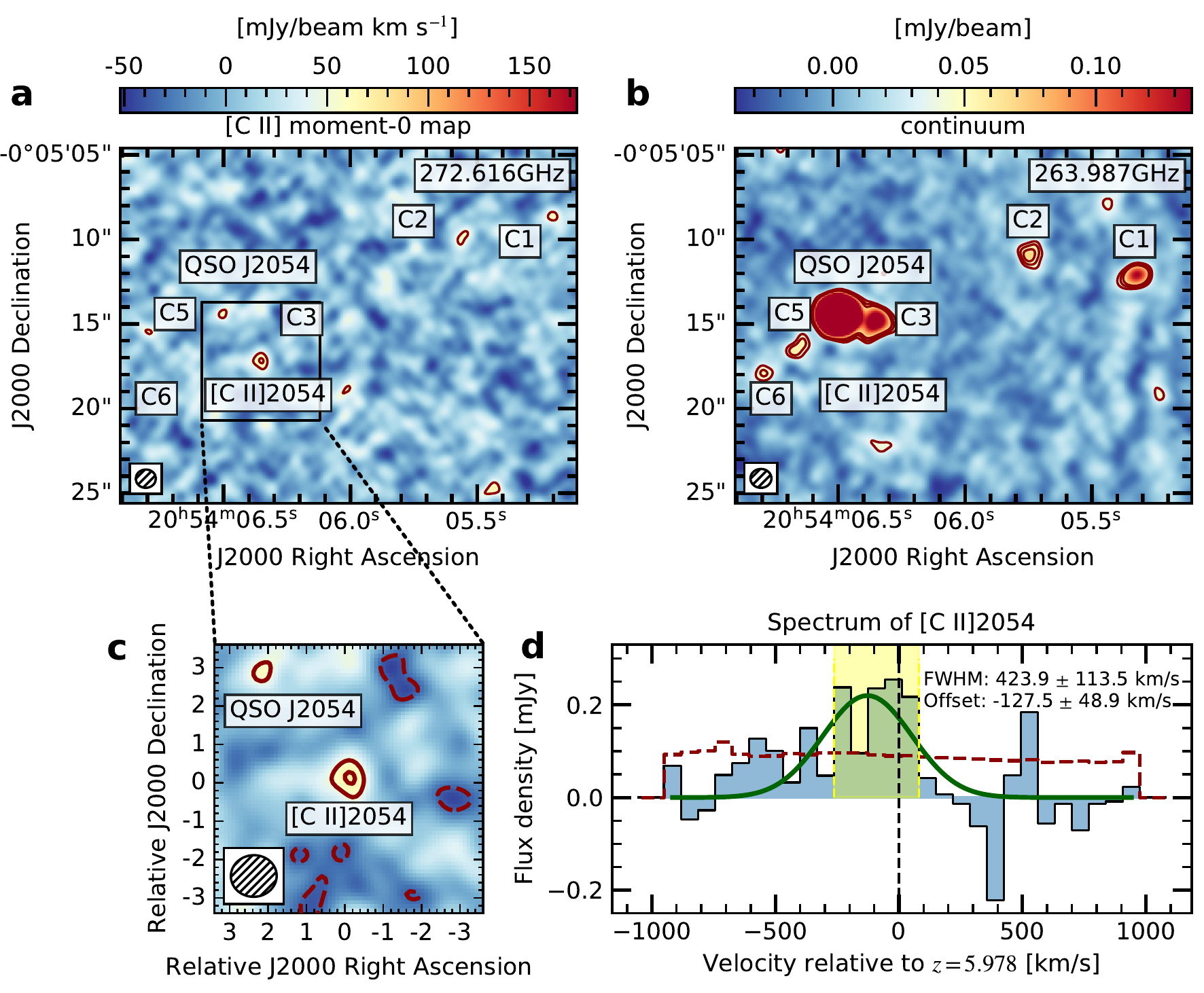}
\end{figure}
\noindent {\bf Fig. 1. ALMA observations.} 
{\bf a,} \CII\ moment-0 map. The outer contour is at 3$\sigma$ level, with contours in steps of 1-$\sigma$. The 1-$\sigma$ is $0.0177\ \rm{Jy~beam^{-1}~km~s^{-1}}$, calculated using the pixel-to-pixel standard deviation. The QSO appears in the continuum-subtracted image because of a water line of ${\rm H_2O}$ with the molecular transition of ${\rm J=3_{2,2}-3_{1,3}}$ at $z=6.04$. {\bf b,} Continuum map. Contours are drawn at 3$\sigma$, 4$\sigma$ and 5$\sigma$. The 1-$\sigma$ noise is 12.4 $\mu$Jy beam$^{-1}$. Six continuum sources within the primary beam can be seen. C4 is the QSO~J2054. Color bars on the top of images represent the integrated flux and flux density respectively. {\bf c,} Zoom-in \CII\ moment-0 map. The sizes of the synthesized beams are plotted in the bottom-left of these panel. The dashed contour is at -2$\sigma$ level. {\bf d,} Spectrum of \CII2054. The red dashed line shows the 1-$\sigma$ rms noise. The \CII\ moment-0 map is collapsed based on the emission range shown by the yellow-shaded region. The {\color{black} black dashed line represents velocity that is relative to the \OI\ redshift $z=5.978$}. The darkgreen line shows a {\color{black} single-Gaussian} model fit to the data. The integral \CII\ flux is $0.0758\pm 0.0177$ Jy km s$^{-1}$.

\newpage
\begin{figure}[!t]
\includegraphics[width=0.95\textwidth]{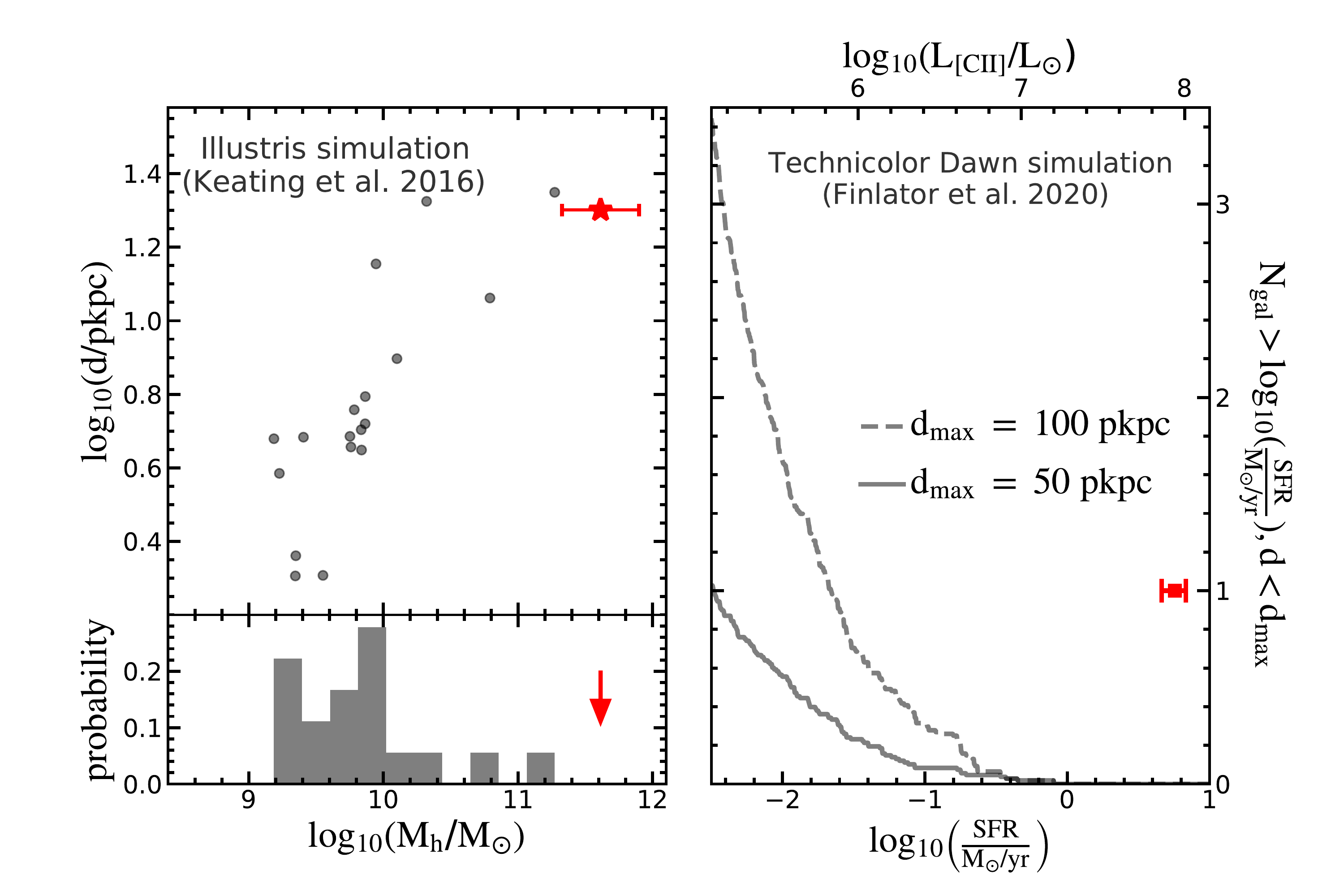}
\end{figure}
\noindent {\bf Fig. 2. Comparison between simulations and observations.} 
\textit{\bf Left Upper:} The relationship between projected impact parameter and halo mass of strong \OI\ absorbers in the Illustris simulation\cite{Keating2016}. The \OI\ absorbers have the rest-frame equivalent width of $0.12\pm0.05$ \AA\ (consistent with observations). Red star represents \CII2054. {\color{black}The error bar shows $1-\sigma$ uncertainty of the estimated halo mass.} {\bf Left Bottom:} Halo mass distribution. Red arrow shows that the host halo mass of \CII2054 is one order of magnitude larger than the median value predicted by Illustris simulations. \textit{\bf Right: }The number of galaxies that cluster about strong \OI\ absorbers at $z=6$ is predicted to vary with SFR. The solid and dashed curves are biased star formation rate functions \cite{Finlator2020}. They are generated by accumulating catalogs of all galaxies that fall within 50 or 100 pkpc of \OI\ absorbers with REW $>0.12$\AA\ ($N_{\rm \OI}>10^{14.2}\mathrm{cm}^{-2}$). The predicted luminosity of \OI\ associated \CII\ emitter is roughly two orders of magnitude fainter than \CII2054.

\newpage
\begin{figure}[!t]
\centering
\includegraphics[width=0.75\textwidth]{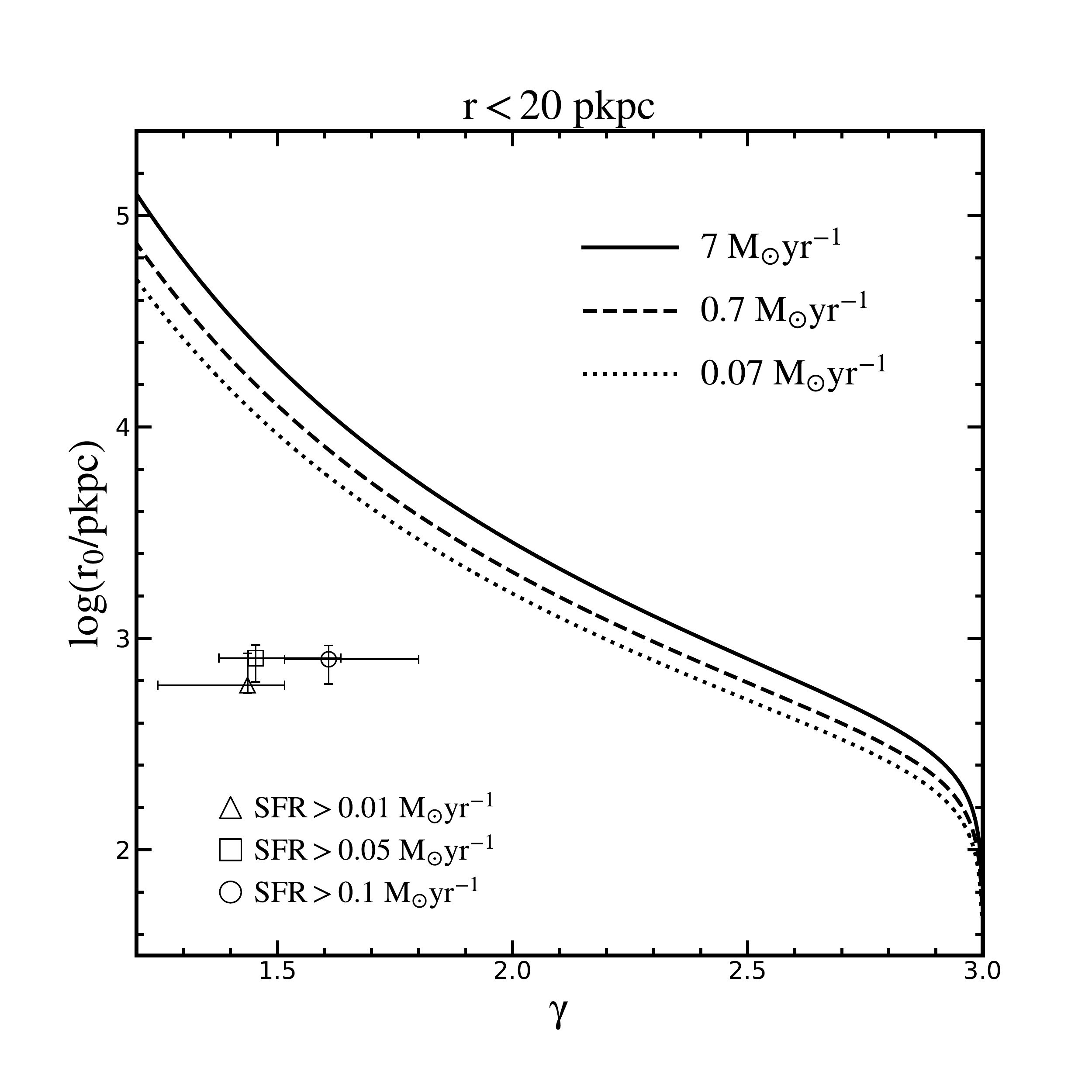}
\label{fig3}
\end{figure}
\noindent {\bf Fig. 3. Comparison of cross-correlation function between simulations and observations.} 
We assume the galaxy-absorber cross-correlation as $\xi(r)=(r/r_0)^{-\gamma}$. This function expresses the fractional excess number density of galaxies located at a distance $r$ from an absorber in terms of the correlation length $r_0$ and a power-law slope $\gamma$. Shaped points indicate cross-correlation functions predicted by Technicolor Dawn for galaxies clustered about synthetic OI absorbers with REW $\ge 0.12$\AA. Error bars are added by assuming Poisson fluctuations in the number of simulated host galaxies. The solid, dashed and dotted line shows that predict an average of one galaxy of SFR $\ge 7,\ 0.7$ and $\ 0.07\ {\rm M_{\odot}/ yr}$,  within 20.0 pkpc of an absorber at $z=5.9$, respectively. Comparing with shaped dots, the predicted $r_0$ is $\sim10\times$ too low to explain the observations. 

\newpage
\begin{figure}[!t]
\centering
\includegraphics[width=0.9\textwidth]{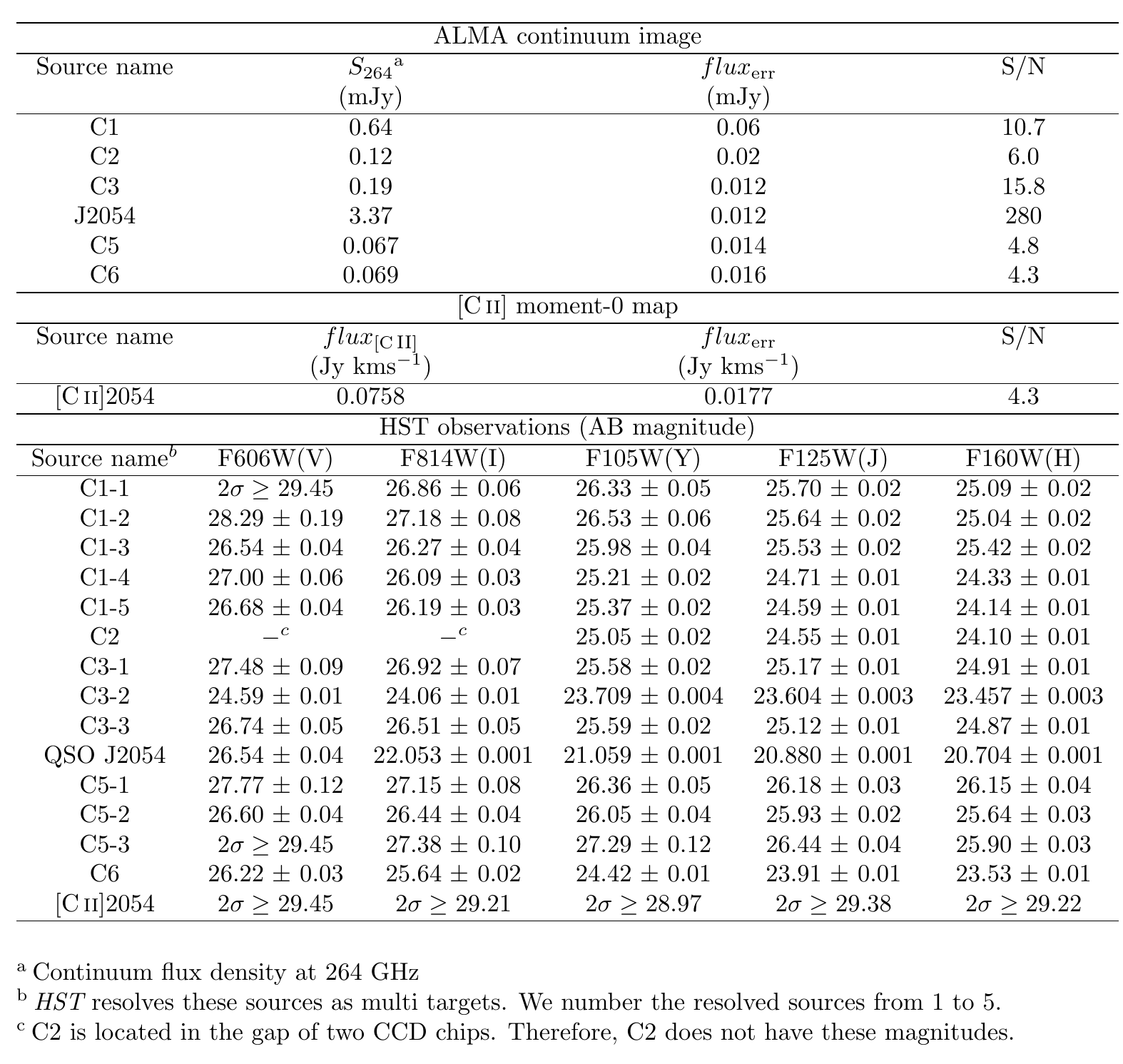}
\label{table1}
\end{figure}
\noindent {\bf Table~1. Photometric Results of sources in J2054 field.} In our ALMA observations, for each source, 2-$\sigma$ contour areas are regarded as emission regions. We use these emission regions as the photometric aperture for resolved sources, while for unresolved sources, the peak surface brightness represents the total flux. For HST observations, we used a uniform aperture with 0.6\arcsec diameter and photometry are measured in the PSF-matched multi-band images (Methods).

\newpage
\begin{methods}
{\color{black}
\subsection{Motivation of this observational project.} 
The observational program was inspired by the successful detection of two host galaxies of the damped Ly$\alpha$ absorbers (DLAs) in 2017 \cite{Neeleman2017}. One of the galaxies has the unexpected SFR of $110\pm10 \msolyr$ {\color{black}[ref. \cite{Fumagalli2015}]}. This DLA is also associated with a strong OI absorber. The OI absorbers picked in this work is among the highest equivalent width at $z>5$ and is probably associated with a DLA or sub-DLA \cite{Becker2011, Becker2019}. These observations were then designed to constrain the galaxy counterparts associated with \OI\ absorbers at $z\approx6$. Based on the \OI\ absorber survey, six out of 17 sightlines contain strong absorbers that can be detected in the QSO spectra. There are 10 absorbers in the six sightlines. Further, two out of 10 \OI\ absorbers are strong enough to have the associated SiII $\lambda$1260/1304 and CII $\lambda$1334 absorption, indicating that they could also be associated with DLAs or sub-DLAs located in the HI Gunn-Peterson trough. Assuming each QSO sightline can probe 500 comoving Mpc, then this strong \OI\ absorber is selected from a survey of $\approx1$ comoving Gpc, indicating that the selected strong \OI\ is rare.
}

\subsection{ALMA observations and data reduction.}

There are a few advantages of using ALMA to probe the source galaxies associated with OI absorber at $z\approx6$. Firstly, the sub-mm \CII\ 158{\rm $\mu$m} line is one of the strongest far-infrared lines from galaxies \cite{De_Looze2011, Carilli2013}. Secondly, the background QSO is less dominant in emission at sub-millimeter regime. Thirdly, at this redshift, \CII\ emission is shifted within ALMA band-6, which has a high throughput and is thus efficient for emission line detections. Finally, cosmological simulations suggest that \OI\ absorbers have a relatively small halo, with predicted impact parameters less than $10$\arcsec, perfectly within the ALMA field of view\cite{Keating2016}. 

{\color{black} The ALMA observations used four 1.875 GHz spectral windows (SPWs) and were obtained in the Time-Division-Mode (TDM).} Each SPW was divided into 43 channels. Two of the SPWs were centered on the \CII\ emission at the redshift of \OI\ absorber and QSO respectively. The remaining two SPWs were used to obtain a continuum image of the field. The on-source time was 2.3 hours. The data were reduced using the CASA 5.4.0 pipeline\cite{CASA2007} following the standard calibration procedures. To obtain the \CII\ emission-line data cube, we subtracted the continuum via the CASA task \emph{uvcontsub} on the line free frequencies. After having the calibrated and continuum-subtracted {dataset}, we generated the image cubes using the task \emph{tclean} with natural weighting to reach the highest sensitivity. The \CII\ intensity map (i.e. moment-0 map) was generated from the \CII\ emission-line channels based on the task \emph{immoments} within CASA. We define the emission-line channels as flux greater than 1-$\sigma$ of the noise continuously. Then, the \CII\ emission FWHM of $\approx400\ {\rm km/s}$ is consistent with a galaxy with SFR of $\sim10$ $\msolyr$. The flux scale was calibrated by observing J2148+0657, whereas the phase stability was checked by observing J2101+0341. Absolute flux uncertainties are expected to be less than 10\%. 

The final data cubes and continuum images were obtained using natural weighting in order to maximize sensitivity. This resulted in a beam size of $1.2$\arcsec $\times1.1$\arcsec and a 1-$\sigma$ background standard deviation of 12.4 $\mu$Jy beam$^{-1}$ for the continuum image. A {\color{black} moment-0 map} image was also generated centered on the redshifted \CII\ emission line. For the {\color{black} moment-0 map}, the emission line channels were collapsed resulting in a beam size of $1.2$\arcsec $\times1.1$\arcsec and a 1-$\sigma$ sensitivity of $0.0177\ \rm{Jy~beam^{-1}~km~s^{-1}}$. For each source, 2-$\sigma$ contour areas are regarded as emission regions. We use these emission regions as the photometric aperture for resolved sources. While for unresolved sources, the peak surface brightness represents the total flux. \CII2054 is unresolved with a de-convolved size of 0.95$'' \times$ 0.74$''$. The photometry results are shown in Table~1. To further discuss the credibility of the detection of \CII2054, we checked signals on the binned channel maps, XX and YY polarization-correlation maps, and three individual exposures. Both XX and YY polarization-correlation maps and all three individual exposures show positive signals. The details are present in the next section. 

\subsection{Details of reliability of \CII2054 detection.}

In this section, we did three tests to study the robustness of the detection of \CII2054. Firstly, we searched for signals in the whole datacube. Each moment-0 image has a width of 344 km s$^{-1}$ in velocity, ranging the whole channel from -641.3 to +734.4 km s$^{-1}$ around the redshift of \CII 2054. {\color{black}None of any other sources having either a signal higher than the $4\sigma$ level or negative signal lower than $-4\sigma$. }Results are shown in the Extended Data Fig. 1 and 2. Note that the QSO is also appear in this continuum-subtracted image because of a serendipitous water line at this frequency. The probability of finding a $4.3$-$\sigma$ source caused by random fluctuation is small. 

Then, we split the XX and YY correlation based on setting the parameter {\tt stokes} in \emph{tclean} to 'XX' and 'YY' respectively. The intensity maps were generated from the same \CII\ emission line channels as mentioned above. The results are shown in Extended Data Fig. 3. The upper panels show the XX intensity map and the corresponding 1-D spectrum. Meanwhile, the lower row shows the ‘YY’ results. The measured integral fluxes are $0.0768\ \pm\ 0.0247$ Jy km s$^{-1}$ and $0.0667\ \pm\ 0.0254$ Jy km s$^{-1}$. The S/N of \CII2054 in XX and YY maps are $3.1$ and $2.6$, respectively, close to expected values of $\frac{4.3}{\sqrt{2}}$. 

Further, we split different exposures based on adjusting the parameter {\tt vis} in \emph{tclean}. The intensity maps are also obtained from the same emission-line channels. Results are shown in Extended Data Fig. 4. The measured integral fluxes are $0.0935\pm0.0319$ Jy km s$^{-1}$, $0.0786\pm0.0335$ Jy km s$^{-1}$, and $0.0490\pm0.0269$ Jy km s$^{-1}$. The S/N of \CII 2054 in the three individual exposures are $2.9, 2.3, 1.8$ respectively, close to the expected value of $\frac{4.3}{\sqrt{3}}$. Thus, none of the evidences support that \CII2054 is a false-positive.

\subsection{Target selection criteria and Source robustness.}

According to our ALMA proposal, our detection limit is down to the typical $L^*$ galaxies at the $\sim4$-$\sigma$ level. These galaxies have the stellar mass of $\sim10^9\ M_\odot$ and the typical halo mass on the order of $\sim10^{11}\ M_\odot$. We expect to marginally detect the $L^*$ galaxies at $\sim4$-$\sigma$ level. Then, according to this expectation, we design the R.M.S. of ALMA observations. Thus, we design our selection criteria based on the marginal detection of a typical $L^*$ galaxy at $z\approx6$. The following three selection criteria are employed: 

Firstly, the \CII\ emitter should have a velocity offset within $\pm150$ km s$^{-1}$ from the \OI\ absorber. This is because that high-$z$ star-forming galaxies with stellar mass of $\sim10^{9}\  {\rm M_{\odot}}$ exhibit outflow velocities ranging $\lesssim$ 150 km s$^{-1}$ {\color{black}[ref. \cite{Steidel2010,Keating2016,Diaz2020}]} at $z=2-6$. Note that our detection limit corresponding to a derived stellar mass of $5\times10^8\ {\rm M_{\odot}}$, residing in a lower or comparable stellar mass with the comparison sample. Further, the projected separation could make the line-of-sight velocity component further lower comparing to the outflow velocity measured along the galaxy sightline \cite{Steidel2010}. 
Therefore, it is safe to constrain the outflow velocity, i.e., the velocity offset between the galaxy and absorber, within $-150 - 150$ km s$^{-1}$. 

Secondly, the projected impact parameter between the source galaxy and the strong \OI\ absorbers have to be smaller than $4.0''$ at $z\approx6$. This is because that strongly metal-enriched gas should be the circumgalactic medium (CGM) and located within the dark matter (DM) halo of the host galaxy. At $z\approx6$, simulations suggest strong OI absorbers with rest-frame equivalent width (REW) equal or greater than 0.1\AA\ have the projected impact parameter within $\le R_{\rm{vir}}$ of the DM halo according to hydro-dynamical simulations (e.g., ref\cite{Oppenheimer2009,Keating2016,Finlator2013,Finlator2020}). {\color{black} Additionally, according to all these simulations, the actual projected impact parameter is within $R\le 0.7\times R_{\rm{vir}}$.} Further, based on our designed halo mass limit of $10^{11} {\rm M}_{\odot}$ (note our detected \CII2054 have the derived halo mass of $4\times10^{11}\ M_\odot$), corresponding to the Virial radius of $20-30$ kpc at $z\approx6$. Thus, when we searching for $\sim L^*$ source galaxies in the field, we set the impact parameter limit of the detection of a $\sim L^*$ galaxy to be  $4.0''$. Note this is also a safe constraint because the projection effect may make the actual separation greater than the limit. 

Thirdly, according to the \CII\ line width analysis, galaxies with the \CII\ luminosity of $L_{\rm{[CII]}}\sim 10^{8}\ L_\odot$ (stellar mass of $10^9\ M_\odot$) have line widths of \CII\ greater than $250$ km s$^{-1}$. This is according to the FWHM measurement from various previous literatures \cite{Fujimoto2019,Bethermin2020,Loiacono2021}. Thus, we search for galaxies with the FWHM of \CII\ greater than 250 km s$^{-1}$. Note the line width criteria may be helpful for further eliminate the false-positive sources generated by noise. Correlating noise may be more frequently generating narrow ``line", because the peak of the noise could influence adjacent channels (also see Extended Data Fig. 5). 

Having all of these three criteria, we searched for emitters from the continuum-subtracted datacube using the algorithm called "FINDCLUMPS" \cite{Walter2016,Decarli2020}. {\color{black}Firstly, we perform floating averages of a given number of channels with different window sizes (e.g. 3-,4-, and 5-channel windows) at different frequencies. After having these maps, the R.M.S. are estimated based on the pixel-to-pixel standard deviation in every averaged map. Then, we searched for peaks exceeding a certain S/N threshold using the python task DAOFIND in every maps. } After proceeding source target-finding algorithm, to get an estimate of the probability of finding peaks having $\gtrsim4$-$\sigma$ significance, we adopt the following three approaches. 

(1) Firstly, we check the probability of $\gtrsim 4$-$\sigma$ in our own spectral windows (SPWs). The four SPWs all reside in band-6, with the central frequency of 272.3, 270.5, 256.0, 257.8\ GHz, respectively. Each has a frequency band width of $\sim2$ GHz, corresponding to a total velocity range of $9\times10^3$ km s$^{-1}$. Two out of the four SPWs were set to observe the \CII\ emission at the redshift of \OI\ absorber and the background QSO respectively. The remaining two SPWs were used to obtain a continuum image of the field. We performed the line-search algorithm on the four SPWs and found that 14 sources in the whole field with S/N $\gtrsim 4\sigma$ (including \CII2054). Excluding two lines (water emission line and QSO \CII\ emission) caused by the background QSO, the number of sources having S/N $\gtrsim 4\sigma$ is 12. For each SPW, we have an effective searching area of $\approx23"$, corresponding to a primary beam limit of 0.01 in the edge of the field of view. Note our selection criteria have a searching area of $\pi\times4''^2 \sim50.2 \ {\rm arcsec^2}$, the velocity range within $\pm150$ km s$^{-1}$, and the linewidth of $\gtrsim 250$ km s$^{-1}$. {\color{black} Scaling it to our selection criteria, we calculate the expected number of sources having $\frac{S}{N}\gtrsim4$-$\sigma$} to be \begin{equation}
    n = 12 \times \frac{\pi\times4''^2\ {\rm arcsec}^2}{\pi\times23''^2\ {\rm arcsec}^2} \times \frac{300\ {\rm km\ s}^{-1}}{9\times10^{3}\ {\rm km\ s}^{-1}}\times \frac{6}{12} = 0.60\%.
\end{equation} 
where we find that six out of 12 emitters have a line width of $\ge 250$ km s$^{-1}$. 

{\color{black}Therefore, using our own survey of four SPWs in band-6, we found the probability of randomly finding $\gtrsim4$-$\sigma$ source is also $0.6\%$ based on our target selection criteria and assuming a Poisson distribution.} Note that the probability calculated from our own survey could be regarded as the most reliable estimate of the probability of $\gtrsim4$-$\sigma$ sources, since the noise properties is identical between \CII2054 and other $\gtrsim4$-$\sigma$ sources.

(2) Further, we estimate the probability finding $\gtrsim 4$-$\sigma$  sources in random-field observations by a source-matching method. Aravena et al. (2016) (ref. \cite{Aravena2016}) searched for \CII\ emitters from the ASPECS-pilot datacubes. In their survey, they searched for line emitters from $\sim1\ {\rm arcmin^2}$ sky area and a $\sim60$ GHz frequency range, with the lower frequency limit of 212 GHz and higher frequency limit of 272 GHz. This frequency range corresponds to a wavelength range of $3.1\times 10^{6}$ \AA, and a velocity range of $\Delta v = 9\times10^4$ km s$^{-1}$. Then, they matched the line-emitter catalog and optical dropout galaxies by constraining the offset within 1$''$ from the whole frequency range. They found that 12 out of 50 optical dropout galaxies have $\gtrsim4$-$\sigma$ line-emitter counterparts. {\color{black}Using results from ref.\cite{Aravena2016} , the expected number of finding targets having $\gtrsim4$-$\sigma$ is} 
\begin{equation} 
n = \frac{12}{50}\times\frac{\pi\times4''^2\ {\rm arcsec}^2}{\pi\times1''^2\ \ {\rm arcsec}^2}\times \frac{300\ {\rm km\ s}^{-1}}{9\times10^4\ {\rm km\ s^{-1}}}\times\frac{6}{12}=0.64\%. 
\end{equation}
where $\frac{6}{12}$ represents the fraction of sources with FWHM $\ge 250$ km s$^{-1}$ to the total number of sources with $\gtrsim4$-$\sigma$ (please see Figure 8 in ref. \cite{Aravena2016}). {\color{black}Therefore, using the survey of ref. \cite{Aravena2016}, we found the probability of finding $\gtrsim4$-$\sigma$ source is $0.64\%$ using our target selection criteria (assuming a Poisson distribution). }

(3) Moreover, we estimate the probability finding $\gtrsim 4\sigma$ sources in a blind line survey. We firstly checked this probability in the newly released deep ASPECS survey over a large field of 4.2 arcmin$^2$ {\color{black} [ref. \cite{Gonzalez_Lopez2019, Uzgil2021}]}. Using the source detection algorithm \cite{Walter2016,Decarli2020}, we searched for line emitters from the $\sim4.2\ {\rm arcmin^2}$ sky area and a $\sim60$ GHz frequency range. The velocity range is larger than that of our selection criteria by a factor of 300. We found that $1209$ targets are identified in the whole 4.2 arcmin$^2$ field (please check  Extended Data Fig. 5.). After scaling it to our searching region ($\pi\times4''^2 \sim50.2 \ {\rm arcsec^2}$), the velocity range within $\pm150$ km s$^{-1}$, and linewidth of $\ge 250$ km s$^{-1}$, {\color{black} the expected number of randomly finding such a target is }
\begin{equation}
    n = 1209\times \frac{300\ {\rm km\ s^{-1}}}{9\times 10^4\ {\rm km\ s^{-1}}}\times\frac{\pi\times4^2\ {\rm arcsec}^2}{4.2\times 3600 \ {\rm arcsec}^2}\times\frac{719}{1209}=0.79\%. 
\end{equation}
where $\frac{719}{1209}$ represents the fraction of sources with FWHM $\ge 250$ km s$^{-1}$ to the total number of sources with $\gtrsim4$-$\sigma$. {\color{black} Therefore, using a large survey over 4.2 arcmin$^2$ area, we find that the probability of randomly finding $\gtrsim4$-$\sigma$ source is $0.79\%$ using our target selection criteria (assuming a Poisson distribution). }

\subsection{The probability of \CII2054 as a CO interloper.}

CO (3-2), CO (4-3), CO (5-4) and CO (6-5) transitions can possibly fall into our observed spectral window, corresponding to the redshift of 0.3, 0.7, 1.1, and 1.5, respectively. Given the measured integral flux is $0.0758 \pm 0.0177\ {\rm Jy\ km\ s^{-1}}$, this corresponds to possible CO line luminosities of $1.8\times 10^{6}$, $1.7\times 10^{7}$, $5.4\times 10^{7}$, and $1.2\times 10^{8}\ {\rm Jy\ km\ s^{-1}\ Mpc^2}$, respectively. After integrating the luminosity function, the mean densities of CO (3-2), CO (4-3), CO (5-4), and CO (6-5) are $2\times 10^{-11}$, $4\times 10^{-11}$, $3\times10^{-11}$ and, $3\times10^{-11}\ {\rm kpc^{-3}}$. Assuming the survey volume is a sphere with a radius of 3.5\arcsec, corresponding to projected distances of 14.4, 24.9, 28.7, and 29.6 pkpc, respectively. After taking all these into account, the probabilities of finding CO in this field are all close to $\sim 10^{-6}$. This indicates that the probability of the CO interloper is extremely low.

\subsection{HST Broadband Imaging.}
For the field surrounding QSO~J2054$-$0005, the {\it HST} archival database has two optical observations F606W(V) and F814W(I) \cite{Marshall2020} and three near-infrared broadband images F105W(Y), F125W(J) and F160W(H) (PID: 15064, PI: Caitlin Casey). These broad filter images are used to estimate the photometric redshifts of targets in this field through the use of the Lyman break drop-out technique \cite{Shapley2003}. Galaxies at $z\approx6$ are expected to drop out of the F606W filter. 

To measure the photometry of the galaxies, we matched the point-spread function (PSF) firstly. We generate PSFs in different filters by stacking stars in the field by using \emph{EPSFBuilder} in Astropy. These stars were matched by using the Gaia archival catalog. After matching PSFs to that of F160W, the measured FWHMs of the PSF in all five bands are 0.23\arcsec (consistent with the FWHM of the PSF in F160W). These results are also consistent with the PSF matching results in CANDELS Multi-wavelength Catalog \cite{Stefanon2017}. Then, we chose the diameter of the photometric apertures as 0.6\arcsec, corresponding to 2.5$\times$ the PSF-matched FWHM. Table~1 shows the HST-broadband photometric results of all the ALMA-detected continuum sources. The HST observations are shown in Extended Data Fig. 6.  

\subsection{\CII\ luminosity, SFR and halo mass estimation } The velocity-integrated \CII\ flux density is measured to be $0.0758\pm 0.0177$ Jy km s$^{-1}$, corresponding to a luminosity of $L_{{\rm \CII}} = (7.0\pm 1.7)\times10^{7}L_{\odot}$. $L_{\rm{\CII}}/ L_{\odot}\ =\ 1.04\times10^{-3}S\Delta v\nu_0(1+z)^{-1}D_L^2$ {\color{black}[ref. \cite{Wang2013}]}, $\nu_0\ =\ 1900.5369$ GHz is the rest frame \CII\ line frequency, $S\Delta v$ is the integrated line flux in Jy km s$^{-1}$, and $D_L$ is the luminosity distance in Mpc. \CII-based SFR, ${\rm SFR_{[CII]}\approx 6.8\pm1.7 M_{\odot}}$ $\rm{yr^{-1}} $ (with 0.3 dex uncertainty)\cite{Schaerer2020}. Comparing with ALPINE survey \cite{Yan2020}, this \CII\ emitter is located at the faint-end. Nevertheless, in a random survey, the probability of finding such a galaxy in this survey volume is $2\times 10^{-5}$. Also, the predicted SFRs of OI associated galaxies are typically lower than $\sim$0.1 $\msolyr$ in these cosmological simulations, we detect a bright galaxy with SFR  of 6.8 $\msolyr$. Therefore, for galaxies associated with strong OI absorber, this detection is exceptional.

Because the star formation rate and stellar mass information are not shown in ref.\cite{Keating2016}, the comparison between the halo mass is conducted for a direct comparison with cosmological simulations. We directly derive it from a \CII\ luminosity-$M_{\rm h}$ relation proposed by simulating galaxies with SFR of 0.01 to 330 $\msolyr$ at $z\sim6$ {\color{black}[ref. \cite{Leung2020}]}. Galaxies with similar SFR as \CII2054 can be reproduced in this simulation. Using the relation between \CII\ luminosity and halo mass in their work (eq. 8), the derived halo mass of \CII2054 is $4.1\times 10^{11} {\rm \msun}$, with the 1-$\sigma$ scattering of 0.5 dex. 

To further confirming the halo mass estimation, we use another method of deriving the halomass from the stellar mass. To do this, we need to derive the stellar mass. We convert the \CII-derived total SFR to the stellar mass\cite{Salmon2015, Speagle2014, Capak2015}, then the predicted stellar mass of \CII2054 is $5.8\times10^{8}$ $M_{\odot}$, with 0.3 dex uncertainty. We further calculated the dynamical mass and molecular gas fractions of $f_{\rm molgas} = \frac{M_{\rm molgas}}{M_{\rm molgas} + M_*}$ to confirm the esitmation of stellar mass. From the spectrum of \CII2054, the \CII\ velocity dispersion ($\sigma_{\rm [CII]}$) is 187 km s$^{-1}$, corresponding to the circular velocity of $v_{\rm cir} = 1.763\sigma_{\rm [CII]}/\sin(i)$. Thus, considering the inclination, the lower limit of $v_{\rm cir}$ is 329.7 km s$^{-1}$. Further, according to the mentioned unresolved size of \CII 2054, the circularized effective radii is $r_e = 4.8\ {\rm kpc}$\cite{Fujimoto2020}. The disk-like gas dynamical mass is estimated from ref.\cite{Dessauges-Zavadsky2020} eq. 4, $\frac{M^{\rm rotation}_{\rm dyn}}{M_{\odot}}  = 1.16\times10^5\left(\frac{v_{\rm cir}}{{\rm km\ s}^{-1}} \right)\left(\frac{2r_e}{\rm kpc}\right)$. Therefore, the derived dynamical mass is $3.7\times 10^{8} {\rm \msun}$, consistent with our derived stellar mass\cite{Capak2015}. Then, using the relation between \CII luminosity and molecular gas mass\cite{Zanella2018}, the derived molecular gas mass ($M^{\rm [CII]}_{\rm molgas}$) is $2.0\times10^{9}\ M_{\odot}$. Thus, the molecular gas fractions is 0.78, consistent with the results from the ALPINE survey\cite{Dessauges-Zavadsky2020}. Then, we can estimate the halo mass from the stellar mass based on Equation 1 of ref.\cite{Ma2018}. The predicted halo mass is about $1.1\times 10^{11}\ M_\odot$. Note the conversion between the stellar mass and halo mass has a scatter of 0.3 dex. In this method, the halo mass is estimated by using three scaling relations: \CII\ luminosity-SFR, SFR-stellar mass, and stellar mass-halo mass. All of these derivations have scatters of $\sim$0.3 dex. Thus, the combined uncertainties is close to 3 times a  0.3 dex scatter. Thus, this estimate is consistent with the value we derived in the last paragraph within the 1-$\sigma$ level.

\subsection{\CII2054 continuum properties.}
No continuum was detected at the position of \CII2054, resulting in a 3-$\sigma$ upper limit of 37 $\mu$Jy~beam$^{-1}$. To convert this continuum flux to a total infrared luminosity (TIR; 8-1000$\mu$m), we estimate the TIR luminosity by fitting a modified blackbody (i.e., $S_{\nu}\sim\nu^{\beta}/(\rm{exp}(h\nu/kT_{\rm{dust}})-1)$) to this single upper limit. We assume \CII2054 has the same continuum properties as a typical Lyman break galaxy at $z\approx6$, and therefore we adopt a dust temperature of $T_{\rm{dust}} = 38$K and emissivity index of $\beta=2.0$ {\color{black}[ref.\cite{Faisst2020}]}. This translates to a 3-$\sigma$ upper limit of $L_{\rm{TIR}}=7.3\times10^{10}$ $L_{\odot}$. Assuming a graybody spectrum\cite{Murphy2011,Izumi2018}, the dust temperature and emissivity index mentioned above, we obtained the TIR-based SFR of $\rm{SFR}_{\rm{TIR}} < 11$ $M_{\odot}\ \rm{yr^{-1}}$. \CII2054 also was not detected in any of the HST broadband imaging using SExtractor\cite{Bertin1996} (Extended Data Fig.~6). From the HST imaging, we determine a 3-$\sigma$ upper limit of the UV continuum at the position of \CII2054 of $3.413\times10^{44}$ erg s$^{-1}$ (F105W band). This corresponding to an upper limit of the UV-based SFR of $\rm{SFR}_{\rm UV} < 15$ $M_{\odot}~\rm{yr^{-1}}$.

\subsection{Continuum sources.}
There are five {\it HST} broad-band observations in this field (F606W(V), F814W(I), F105W(Y), F125W(J), F160W(H)), the photometric results for each source are tabulated in Table~1. To constrain the photometric redshifts of these continuum sources, we use color-color diagrams as shown in Extended Data Fig.~7, and compare the colors of the galaxies with those colors of simulated galaxies. In the {\it HST} image (Extended Data Fig. 6), C1, C3, C5 are resolved as multiple targets. Therefore, we plot the colors of these multiple sources individually. 

For the model galaxies, we use a general star-forming galaxies approach \cite{Ono2018}. Model spectra of different galaxies at different redshifts are generated from the models presented in BC03 \cite{Bruzual2003}. For the galaxy models, we take the metallicity to be $Z = 0.02Z_{\odot}$, with an age of 25 Myr, and we assume a Salpeter initial mass function. We further assume a constant SFR of 1 M$_{\odot}~$yr$^{-1}$, with an exponentially declining star formation history with timescales of $\tau _{\rm SFR}=20$ Gyr. 
To convert these galaxy templates to colors, we firstly apply a dust attenuation \cite{Calzetti2000} with E(B-V) = 0.0 - 0.4. Then, we take into account the effects of IGM attenuation\cite{Inoue2014}.
Then, we convolve the spectral energy distribution with the transmission curves of the five {\it HST} broad band filters to get the colors of the simulated galaxies. In the left panel of Extended Data Fig.~7, the dark blue dots are the low-redshift template galaxies with redshift from $z=2.0-4.0$. The red points represent star-forming galaxies at $z=5.5-6.5$. 
We use the following criteria to select galaxies at $z>4$.
\begin{equation}
    F606W - F814W > 2
\end{equation}
\begin{equation}
    F814W - F105W < 3
\end{equation}
\begin{equation}
    F606W - F814W > 1.5\times(F814W - F105W) + 0.5
\end{equation}
Based on these criteria, we rule out almost all of the continuum sources as high redshift candidates, except for C1-1 and C5-3. In this panel, C1-1 and C5-3 are not detected in the F606W filter, which could indicate that they are V-dropout galaxy candidates. In the right panel of Extended Data Fig. 7, we further study their color using the F814W, F105W and F125W filters, and the following criteria to select $z\approx6$ galaxies:
\begin{equation}
    F814W - F105W > 0.2
\end{equation}
\begin{equation}
    F105W - F125W < 0.4
\end{equation}
\begin{equation}
    F814W - F105W > 1.6\times(F105W - F125W) + 0.36
\end{equation}
With these criteria, C1-1 and C5-3 are both ruled out as galaxies at $z\approx6$. We conclude that, except C2, all of the continuum sources are regarded as low-$z$ foreground galaxies. The redshift of C2 is difficult to constrain without the F606W and F814W imaging.

\subsection{Correlation length and power-law index calculation.} 
In this section, details are shown about the calculations of the correlation length. We adopt the same method displayed in ref.\cite{Finlator2020} $\xi_{\rm g-abs}(r)=\frac{1}{n_0}\frac{\Delta N(r)}{\Delta V} - 1$, where $\Delta N(r)$ representing the mean number of galaxies within a finite spherical shell with the of volume $\Delta V$ located a distance $r$ from an absorber, while $n_0$ being the mean number density of galaxies averaged over the entire simulation volume. Further, this equation reduces as $n_0[1+\xi_{\rm g-abs}(r)] = \frac{\Delta N(r)}{\Delta V}$. To quantify this relation more physically, we assume this correlation obeying a power law by introducing the correlation length ($r_0$) and a power-law index ($\gamma$), i.e. $\xi_{\rm g-abs}(r)= \left(\frac{r}{r_0}\right)^{-\gamma}$. After reducing $\xi_{\rm g-abs}$ in these two equations, we obtained $n_0\left[1+(\frac{r}{r_0})^{-\gamma}\right] = \frac{\Delta N(r)}{\Delta V}$. By integrating this equation over r, we got $4\pi \int_0^rn_0\left[1+(\frac{r}{r_0})^{-\gamma}\right]r^2 dr = N $. Further, we only detect one galaxy in the survey volume. Thus, $N=1$ and $4\pi \int_0^rn_0\left[1+(\frac{r}{r_0})^{-\gamma}\right]r^2 dr = 1$. Finally, we obtain $\log(r_0) = \frac{\log[A(\gamma)]}{\gamma} + \log(r)$, where $A\equiv\left(\frac{1}{4\pi n_0} - \frac{r^3}{3}\right)\times \frac{3-\gamma}{r^3}$.

To compare with observations, the number of galaxies that are brighter than \CII2054 within the observed projection impact parameter ($r = 20{\rm pkpc}$) are calculated. The mean number density ($n_0$) of \CII\ emitter at this redshift was calculated by integrating the \CII\ luminosity function\cite{Yan2020}. Thus, the mean density of \CII\ emitter at $z\sim6$ is $1.4 \times 10^{-3}\ {\rm cMpc^{-3}}$. Assuming a spherical volume with the radius of 20 pkpc, this volume containing one such bright galaxy has a mean overdensity of $\Delta = 5\times 10^4$. Thus, with a fixed $r$, the $r_0$ and $\gamma$ relation can be obtained and shown in Figure 3.

\subsection{Comparison between observations and different simulations.} 
We compare the measured impact parameter and the derived halo mass of \CII2054 and that of other three simulations, i.e., Sherwood, HVEL, and FAST simulations. {\color{black} The comparisons are shown in the Extended Data Fig. 8. }These absorbers are selected with REWs of $0.12\pm0.05$\AA. We find that, as the mass of host halo increases, absorbers can be ejected to increasing distances from the halo. These absorbers are located closest to the host halos with typical masses of $10^{10} M_{\odot}$. However, the derived halo mass of \CII2054 is one order of magnitude larger than the median value predicted by all of these simulations. Our results suggest that low-mass systems in these simulations are overly efficient at producing metal absorbers. {\color{black} Further, Our observations strongly constrain all of these simulations.}

\subsection{The results of the second QSO feld.} 
One may note that the results reported here are only a part of our ALMA  observational program. Actually, three fields have been proposed and two out of the three fields have been finished. The analysis of the second field was finished recently. Nevertheless, the third field have not been observed yet. With the preliminary analysis, none detection have been yielded in the second QSO field. Thus, if including the other field, then our survey volume is larger by a factor of two. The probability of randomly finding a galaxy in our survey volume will be $4\times 10^{-5}$, still much smaller than the unity.

\subsection{Data Availability}
Both ALMA and HST data used in this work have been public available. The data reported in this paper are available though the ALMA archive: (http://almascience.eso.org/aq/) with project code: 2017.1.01088.S, the HST (https://archive.stsci.edu/hst/) with project code: 15410 and 12974. Other data are available from the corresponding author based on reasonable request.

\subsection{Code Availability}

The ALMA data were reduced using the CASA pipeline version 5.4.0, available at https://casa.nrao.edu/casa\_obtaining.shtml.

\newpage
\begin{figure}[!t]
\centering
\includegraphics[width=1\textwidth]{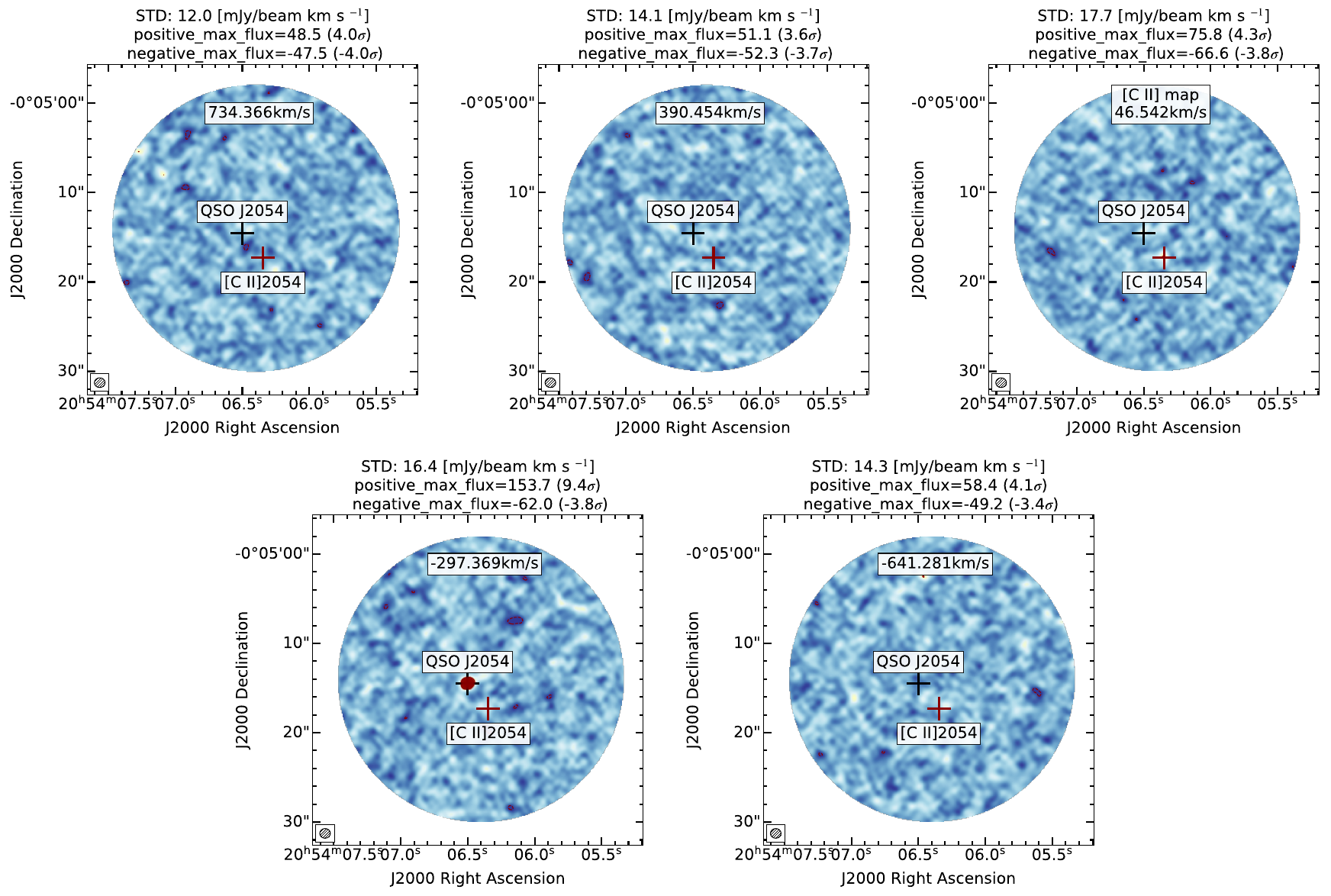}
\end{figure}
\noindent {\bf Extended Data Fig. 1. Channel maps in the whole observed field of view.} 
These channel maps are arranged from $-641.281$ to $+ 734.366$ km/s. The channel width is set as the same as \CII\ intensity map as 344 km/s for each map. The black and dark red cross indicate the position of QSO J2054 and \CII2054, respectively. The sizes of the synthesized beams are demonstrated in the bottom-left of these panel.

\newpage
\begin{figure}[!t]
\centering
\includegraphics[width=1\textwidth]{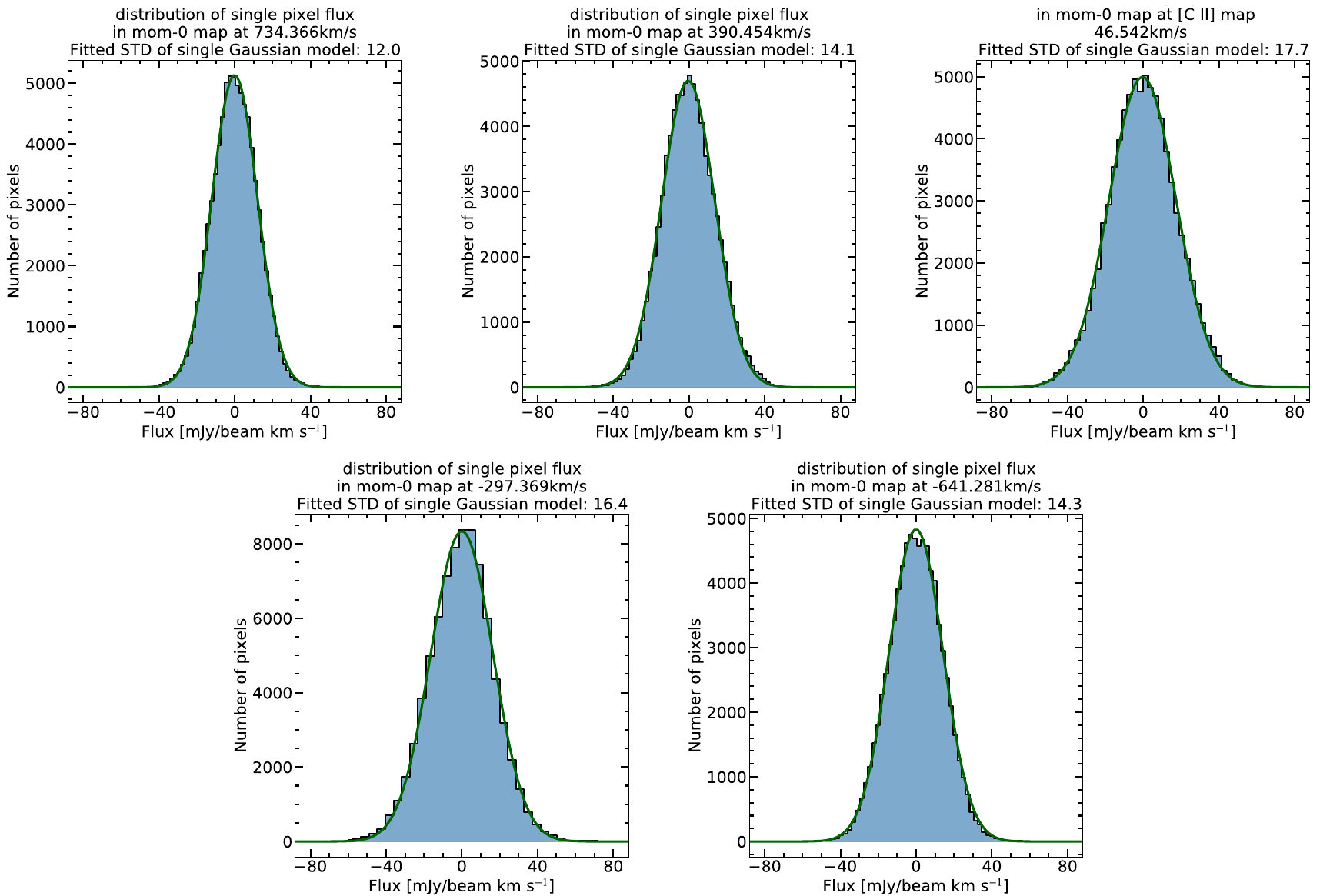}
\end{figure}
\noindent {\bf Extended Data Fig. 2. The detailed pixel flux distributions of different Channel maps.} 
The green lines show the best-fit single-Gaussian models. We regard the single-Gaussian fitted standard deviation (STD) as the noise in these intensity maps.

\newpage
\begin{figure}[!t]
\centering
\includegraphics[width=0.9\textwidth]{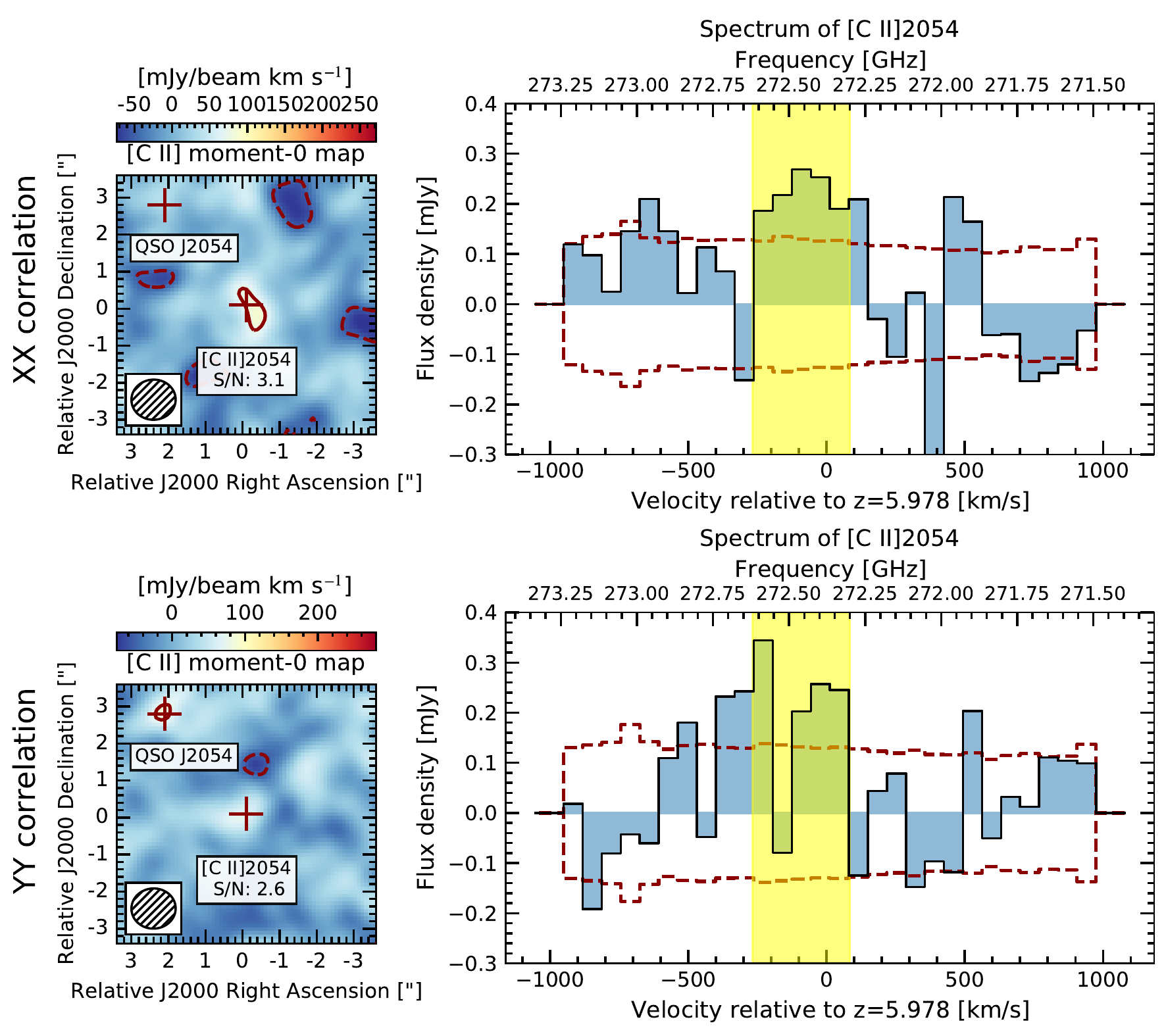}
\end{figure}
\noindent {\bf Extended Data Fig. 3. Different polarization-correlation maps.} 
{\bf Upper panels}, The moment-0 and spectrum under XX polarization. The \CII\ moment-0 maps is collapsed based on the same emission range as Fig.1 and shown by the yellow shaded region. The outer contour is at 3$\sigma$ level, with contours in steps of 1-$\sigma$. The 1-$\sigma$ rms is $2.47\times10^{-2}\rm{Jy~beam^{-1}~km~s^{-1}}$. Dashed lines represent -2$\sigma$-level contours. The integral \CII\ flux is $0.0768\ \pm\ 0.0247$ Jy km s$^{-1}$. {\bf Lower panels,} The resutls under YY polarization. The integral \CII\ flux is $0.0667\ \pm\ 0.0254$ Jy km s$^{-1}$, while the 1-$\sigma$ is $2.54\times10^{-2}\rm{Jy~beam^{-1}~km~s^{-1}}$.

\newpage
\begin{figure}[!t]
\centering
\includegraphics[width=0.7\textwidth]{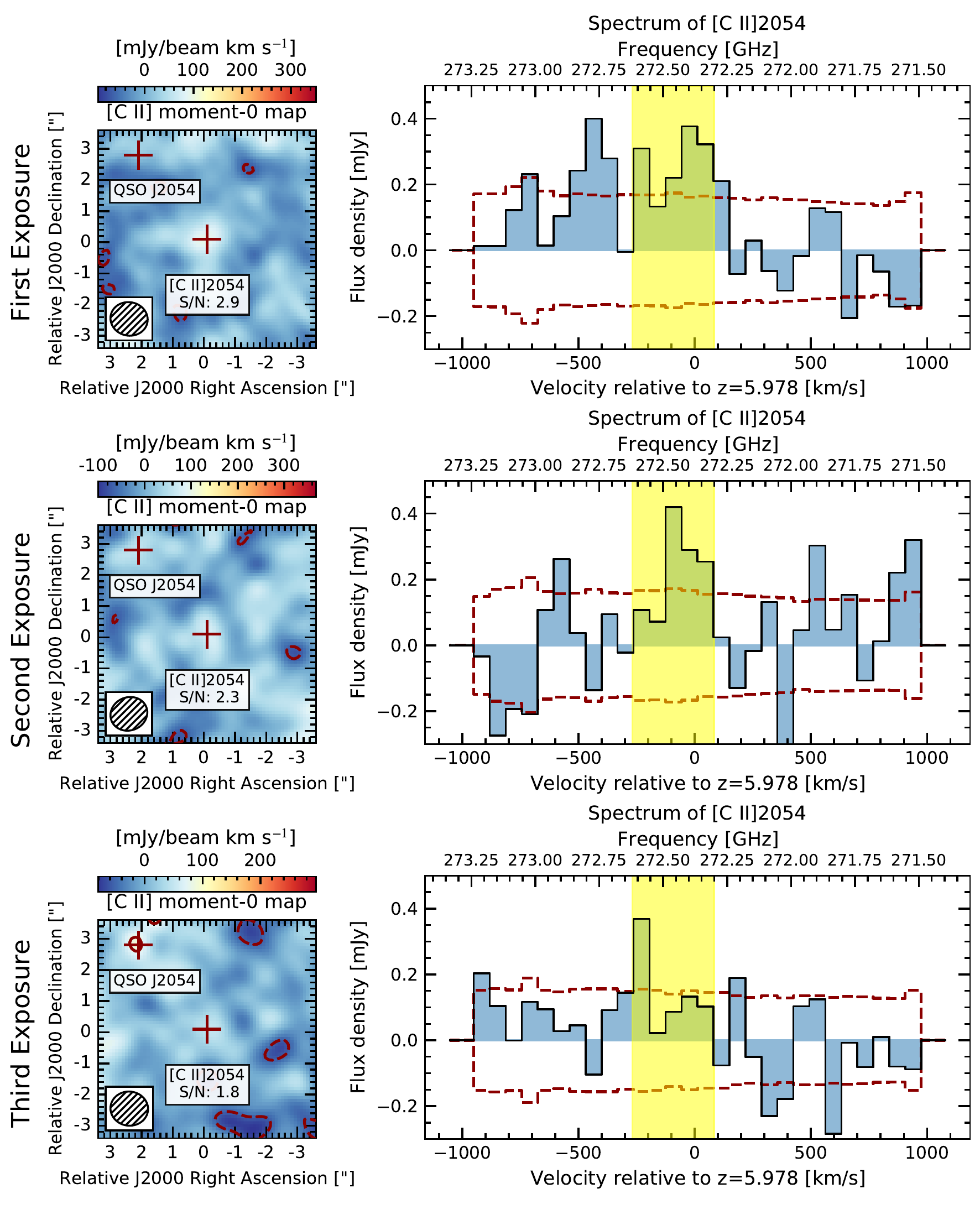}
\end{figure}
\noindent {\bf Extended Data Fig. 4. Multi-exposure observations.} 
{\bf\ Left panel:} The intensity maps of three individual exposures. The integral \CII\ fluxes are $0.0935$, $0.0786$ and $0.0490\rm{\ Jy~beam^{-1}~km~s^{-1}}$. Meanwhile the 1-$\sigma$ standard deviation are $0.0319$, $0.0335$, and $0.0269\rm{\ Jy~beam^{-1}~km~s^{-1}}$, respectively.{\bf\ Right panel:} The corresponding 1-D spectra of \CII 2054 in the different individual exposure shown in the left. The beam sizes are shown in the bottom-left of each mom-0 map. The yellow shaded region shows the emission range that is used to generate the \CII moment-0 maps.

\newpage
\begin{figure}[!t]
\centering
\includegraphics[width=1\textwidth]{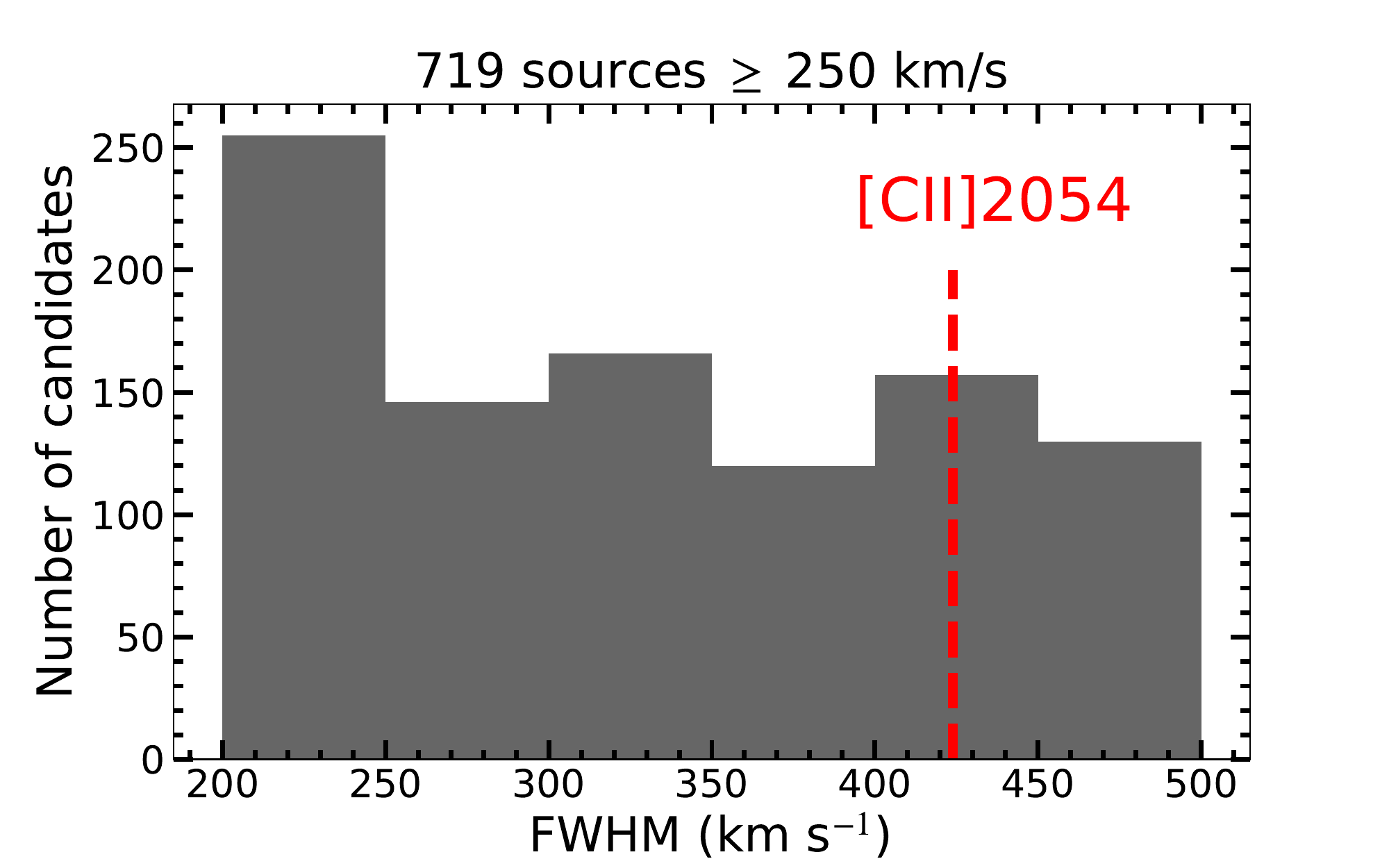}
\end{figure}
\noindent {\bf Extended Data Fig. 5. The number of candidates in the deep ASPECS Band-6 datacube.} 
We found 719 sources having FWHM $\geq250$ km s$^{-1}$ in a 4.2 arcmin$^2$ area of the ASPECS survey. In the figure, sources with Full Width Half Maximum (FWHM) between 200 and 500 km s$^{-1}$ are present. The vertical dashed line represents the FWHM of the single Gaussian fitting of the \CII2054.

\newpage
\begin{figure}[!t]
\centering
\includegraphics[width=0.6\textwidth]{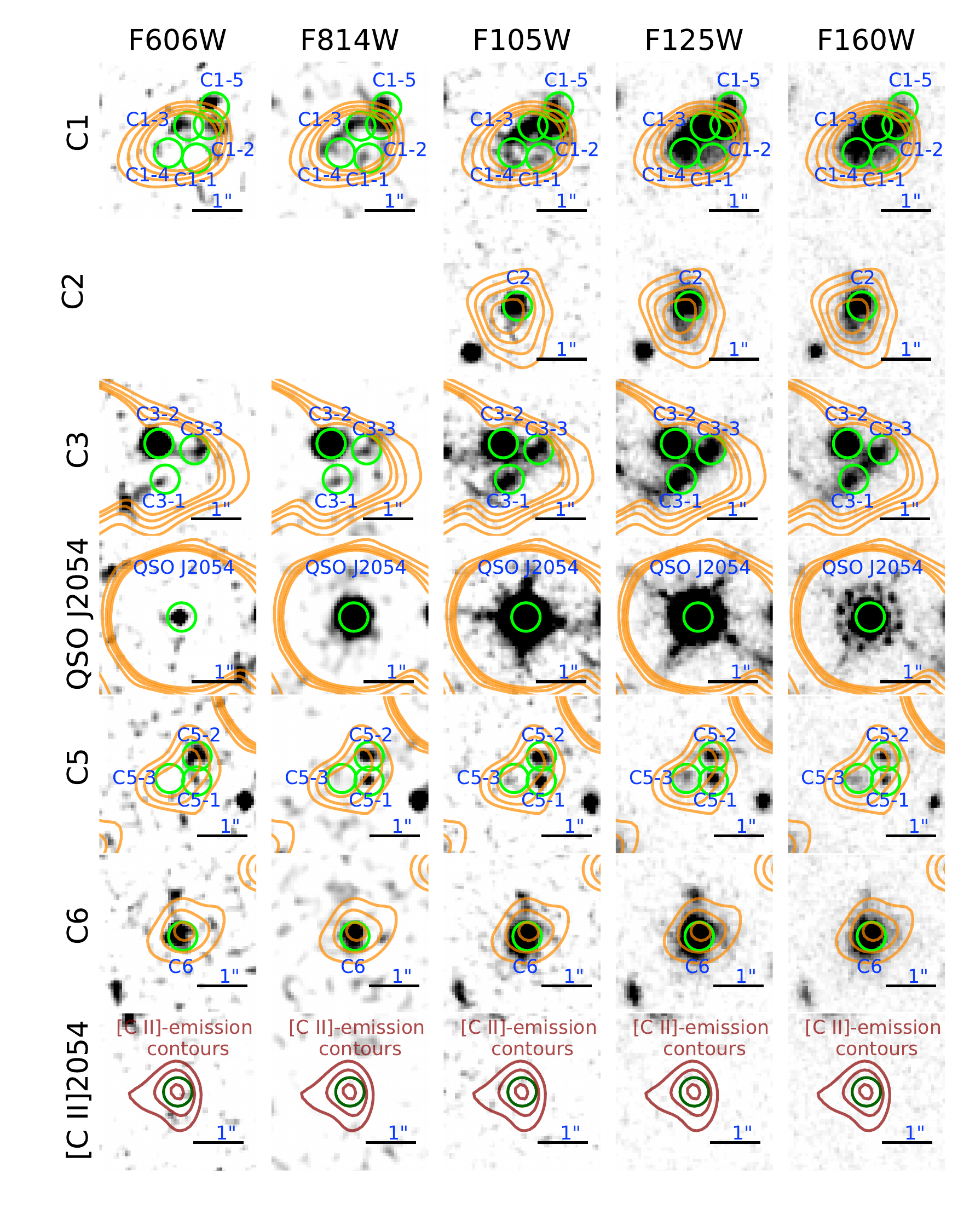}
\end{figure}
\noindent {\bf Extended Data Fig. 6. High-resolution HST broad-band images for five different filters.} 
These images are sorted by the filter central wavelength. From left to right, these images are F606W, F814W, F105W, F125W, and F160W, respectively. Further, different rows represent different continuum sources (as defined in Table 1). The HST photometry is based on apertures with a 0.6 arcsec diameter, as shown by the green and darkgreen circles. The orange ($2-5\sigma$) and darkred contours ($2-4\sigma$) represent the 264GHz-continuum and \CII-emission regions in the ALMA observations, respectively. We find no continuum emission in the \CII-emission region of \CII2054 in all five HST images.

\newpage
\begin{figure}[!t]
\centering
\includegraphics[width=1\textwidth]{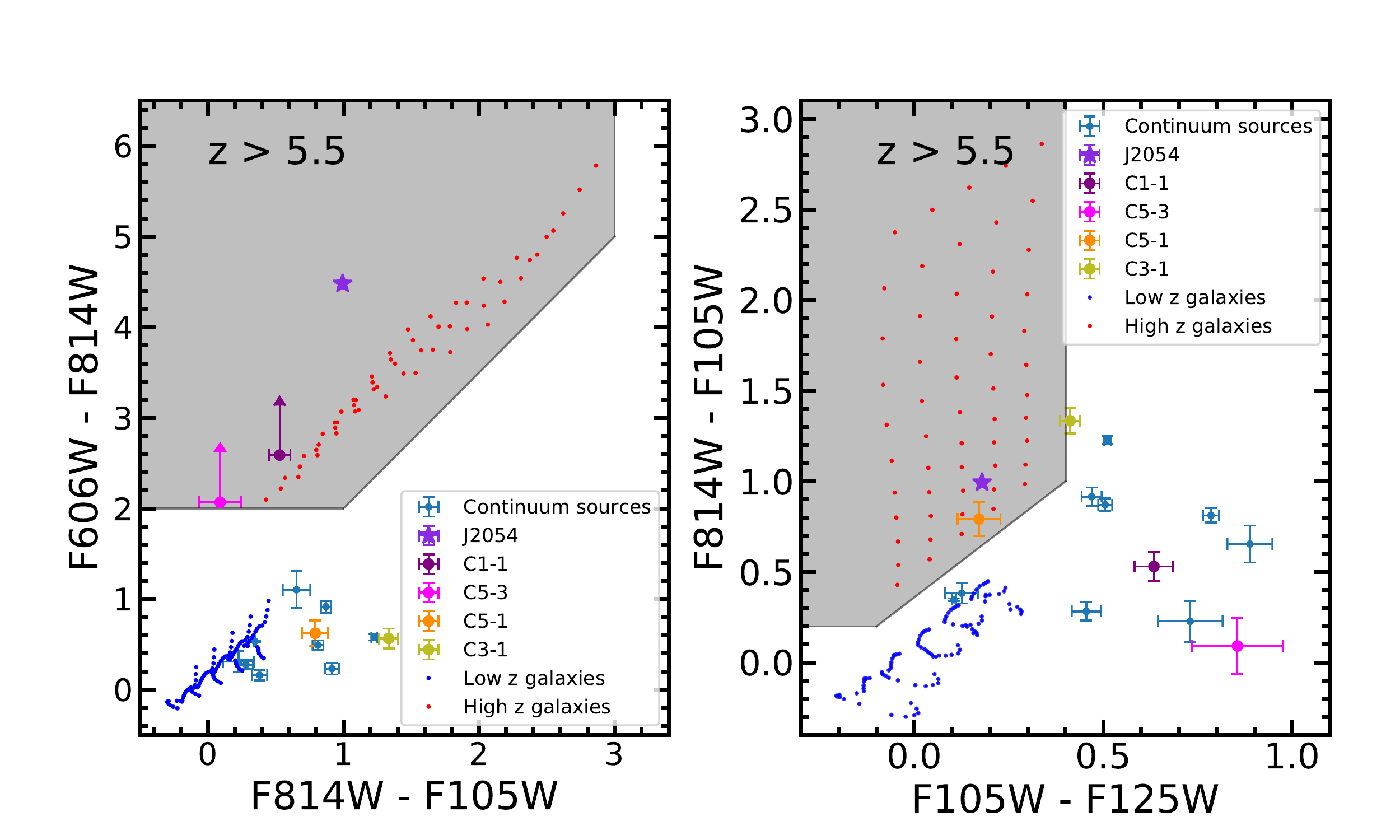}
\end{figure}
\noindent {\bf Extended Data Fig. 7. Color-color diagrams.} 
The color properties of these sources are calculated under the AB magnitudes. The plotted error bars show the propagated uncertainties based on the 1 $\sigma$ errors in magnitudes. The continuum sources are plotted in blue circles. The red and dark blue dots represent the color properties of simulated star-forming galaxies at $z\ >\ 5.5$ and $z\ <\ 4$, respectively. High-redshift selection criteria are based on the distribution of these template galaxies. {\it Left: } $I - Y$ vs. $V - I$ two color diagram. In the left panel, we rule out most continuum sources as high-redshift galaxies, except for C1-1 and C5-3. These two galaxies have the same color properties as the simulated high redshift galaxies, and are plotted as a purple and magenta dot. {\it Right: } $Y - J$ vs. $I - Y$ two color diagram. In the right panel, we rule out C1-1 and C5-3 as high-redshift candidates. Note QSO~J2054-0005 is plotted as a blue-violet star. 

\newpage
\begin{figure}[!t]
\centering
\includegraphics[width=1\textwidth]{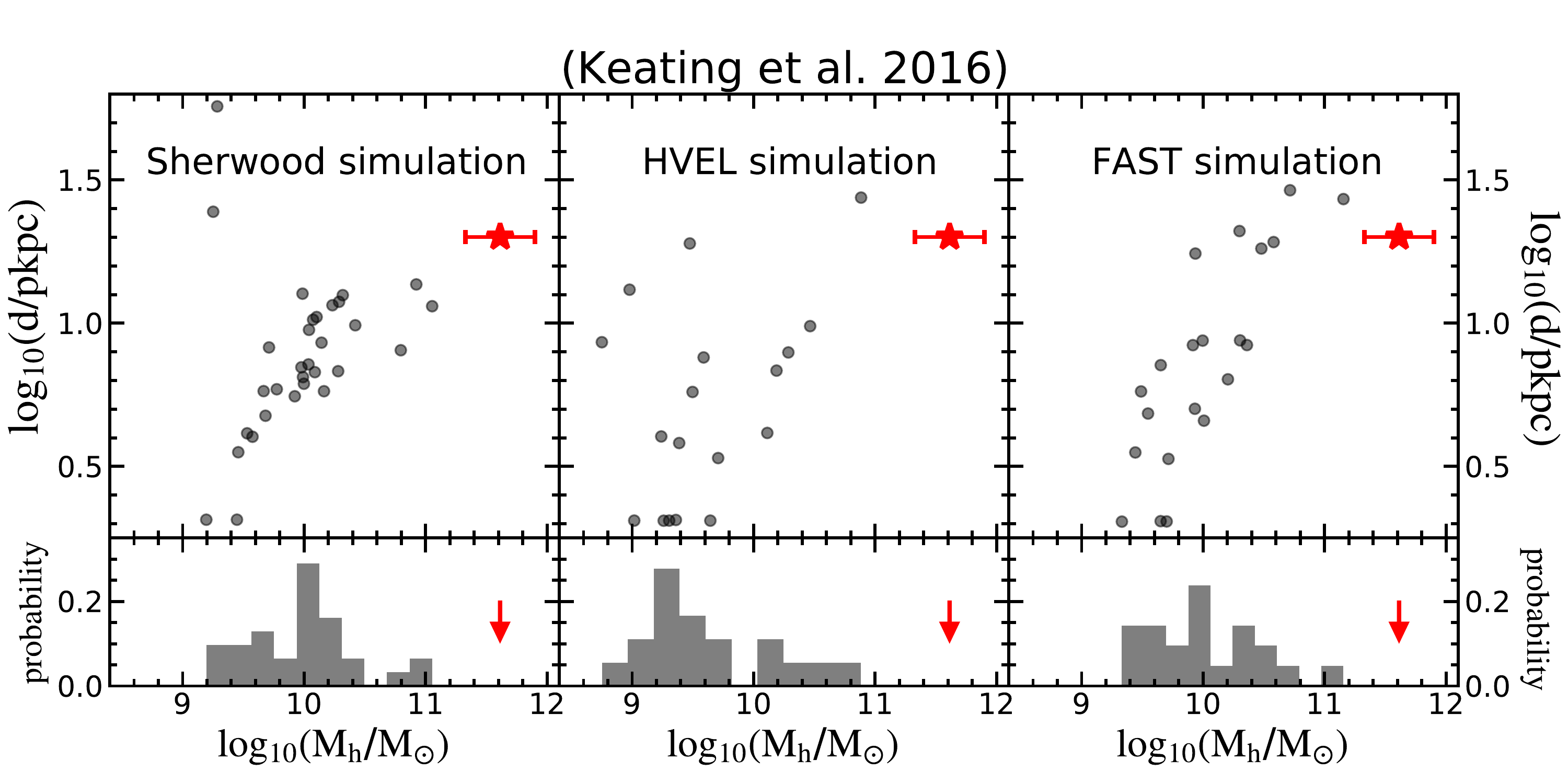}
\end{figure}
\noindent {\bf Extended Data Fig. 8. The relationship between the projected impact parameters and halo masses of strong OI absorbers in different simulations.} 
The halo mass of the \CII2054 is converted directly from the \CII\ luminosity to the halo mass relation \cite{Leung2020}. The error bars represent 1-$\sigma$ uncertainties of the estimated halo mass. In the three top panels, grey dots represent \OI\ absorbers with the REW of $0.12\pm0.05$ \AA\ (consistent with observations).  Meanwhile, Red star represents \CII2054. Bottom panels show the halo mass distribution of different simulations. In the bottom panels, Red arrow shows that the host halo mass of \CII2054 is one order of magnitude larger than the median value predicted by all of these simulations.

\end{methods}

\newpage
\subsection{Acknowledgements}
Z.C. and Y.W. are supported by the National Key R$\&$D Program of China (grant no. 2018YFA0404503) and the National Science Foundation of China (grant no. 12073014). M.N. acknowledges support from European Research Council advanced grant no. 740246 (Cosmic$\_$Gas). F.W. is thankful for support provided by NASA through the NASA Hubble Fellowship grant no. HST-HF2-51448.001-A awarded by the Space Telescope Science Institute, which is operated by the Association of Universities for Research in Astronomy, Inc., under NASA contract NAS5-26555. The National Radio Astronomy Observatory is a facility of the National Science Foundation operated under cooperative agreement by Associated Universities, Inc.

\subsection{Author contributions}
All authors discussed the results and commented on the manuscript. Y.W. and Z.C. led the data reduction, pipeline development, analysis and manuscript writing. Z.C., M.N., K.F. and J.X.P. conceived the project and led the telescope proposal. Z.C. is the principle investigator of the ALMA program (Program ID: 2017.1.01088.S). M.N., K.F. and S.Z. all participated in the analysis and data reduction. R.W. and B.H.C.E. helped with checking of the ALMA data reduction and analysis. X.F., L.C.K., F.W., J.Y., J.F.H. and J.W. all helped significantly with the interpretation and commented on the ALMA proposal and the paper. All authors discussed the results and commented on the manuscript.

\subsection{Competing interests}
The authors declare no competing interests.

\subsection{Correspondence}
Correspondence and requests for materials should be addressed to Z.C. (email: zcai@mail.tsinghua.edu.cn)

\end{document}